\documentclass[sigconf]{acmart}
\usepackage[capitalise,noabbrev]{cleveref}

\copyrightyear{2025} 
\acmYear{2025} 
\setcopyright{cc}
\setcctype{by}
\acmConference[CHI '25]{CHI Conference on Human Factors in Computing Systems}{April 26-May 1, 2025}{Yokohama, Japan}
\acmBooktitle{CHI Conference on Human Factors in Computing Systems (CHI '25), April 26-May 1, 2025, Yokohama, Japan}\acmDOI{10.1145/3706598.3713509}
\acmISBN{979-8-4007-1394-1/25/04}

\begin{document}

\title[People Attribute Purpose to Autonomous Vehicles When Explaining Their Behavior]{People Attribute Purpose to Autonomous Vehicles \\ When Explaining Their Behavior: \\ Insights from Cognitive Science for Explainable AI}

\author{Balint Gyevnar}
\email{balint.gyevnar@ed.ac.uk}
\orcid{0000-0003-0630-4315}
\affiliation{
  \institution{University of Edinburgh}
  \city{Edinburgh}
  \country{United Kingdom}
}
\author{Stephanie Droop}
\email{stephanie.droop@ed.ac.uk}
\affiliation{
  \institution{University of Edinburgh}
  \city{Edinburgh}
  \country{United Kingdom}
}
\author{Tadeg Quillien}
\email{tadeg.quillien@ed.ac.uk}
\affiliation{
  \institution{University of Edinburgh}
  \city{Edinburgh}
  \country{United Kingdom}
}
\author{Shay B. Cohen}
\email{scohen@inf.ed.ac.uk}
\affiliation{
  \institution{University of Edinburgh}
  \city{Edinburgh}
  \country{United Kingdom}
}
\author{Neil R. Bramley}
\email{neil.bramley@ed.ac.uk}
\affiliation{
  \institution{University of Edinburgh}
  \city{Edinburgh}
  \country{United Kingdom}
}
\author{Christopher G. Lucas}
\email{c.lucas@ed.ac.uk}
\affiliation{
  \institution{University of Edinburgh}
  \city{Edinburgh}
  \country{United Kingdom}
}
\author{Stefano V. Albrecht}
\email{s.albrecht@ed.ac.uk}
\affiliation{
  \institution{University of Edinburgh}
  \city{Edinburgh}
  \country{United Kingdom}
}

\renewcommand{\shortauthors}{Gyevnar, et al.}

\begin{abstract}
It is often argued that effective human-centered explainable artificial intelligence (XAI) should resemble human reasoning. However, empirical investigations of how concepts from cognitive science can aid the design of XAI are lacking. Based on insights from cognitive science, we propose a framework of explanatory modes to analyze how people frame explanations, whether mechanistic, teleological, or counterfactual. Using the complex safety-critical domain of autonomous driving, we conduct an experiment consisting of two studies on (i) how people explain the behavior of a vehicle in 14 unique scenarios ($N_1=54$) and (ii) how they perceive these explanations ($N_2=382$), curating the novel Human Explanations for Autonomous Driving Decisions (HEADD) dataset. Our main finding is that participants deem teleological explanations significantly better quality than counterfactual ones, with perceived teleology being the best predictor of perceived quality. Based on our results, we argue that explanatory modes are an important axis of analysis when designing and evaluating XAI and highlight the need for a principled and empirically grounded understanding of the cognitive mechanisms of explanation. The HEADD dataset and our code are available at: \url{https://datashare.ed.ac.uk/handle/10283/8930}.
\end{abstract}

\ccsdesc[500]{Human-centered computing~Empirical studies in HCI}
\ccsdesc[500]{Computing methodologies~Cognitive science}
\ccsdesc[300]{Computing methodologies~Causal reasoning and diagnostics}
\ccsdesc[300]{Computing methodologies~Artificial intelligence}

\keywords{Cognitive science, Explainable AI, Causality, Counterfactuals, Teleology, User study, Autonomous Driving}

\received{12 September 2024}
\received[revised]{10 December 2024}
\received[accepted]{16 January 2025}

\maketitle

\section{Introduction}\label{sec:intro}

The field of explainable AI (XAI) is attracting considerable and growing multi-disciplinary attention. 
Recent years have seen a shift from viewing XAI as a sterile scalpel for dissecting AI models toward using XAI to coordinate knowledge both between expert and non-expert stakeholders and, in a more expansive near-future vision, between natural and artificial agents~\cite{gyevnar2023transparencyGap,paezPragmaticTurnExplainable2019a}.
Cross-disciplinary work in XAI draws on, among others, the social sciences~\cite{miller2019explanation,ehsanWhoXAIHow2024}, psychology~\cite{taylorArtificialCognitionHow2021,yangPsychologicalTheoryExplainability2022, franklin2021blaming}, philosophy~\cite{zednikSolvingBlackBox2021}, and natural language processing~\cite{slackExplainingMachineLearning2023,gyevnar2024cema,madumalGroundedDialogModel2018}.
These collaborations fuel an increasing emphasis on \textit{human-centered XAI}~\cite{ehsanHumanCenteredExplainableAI2020a}.

A central premise of human-centered XAI research is that the design of XAI systems cannot exist in a vacuum but must involve stakeholders from the start~\cite{doshi-velezRigorousScienceInterpretable2017,ehsanHumanCenteredExplainableAI2020a}. This has already resulted in a rich literature that analyzes and categorizes requirements and user preferences~\cite{yuanContextualizingUserPerceptions2023,kimHelpMeHelp2023,nimmoUserCharacteristicsExplainable2024,millerExplainableAIDead2023,liaoConnectingAlgorithmicResearch2022a}.
A common observation throughout these works is that `participants prefer explanations that resemble human reasoning and explanations'~\cite[][p.9]{kimHelpMeHelp2023} and explanations should `align with the cognitive decision-making process that people use when making judgments'~\cite[][p.1]{millerExplainableAIDead2023}.

This `cognitive approach' to designing XAI systems has been heavily popularized by Miller~\cite{miller2019explanation}, such that significant efforts were invested into creating algorithms that provide \textit{causal explanations} (for a comprehensive survey, see \cite{stepinSurveyContrastiveCounterfactual2021}).
Unfortunately, most of this work lacks a robust human-centered motivation for the particular causal framework.
This is apparent not only in the lack of user evaluations used to validate explanation generation methods~\cite{keane2021if} but also in the choice of quantitative metrics, which are ill-suited to assess whether people would give these explanations on their own. 
A major issue here is that work often focuses solely on explanation fidelity (i.e., how faithful an explanation is to the algorithm) as an objective metric, even though different people may understand an explanation differently and value different causal content depending on their context or background~\cite{wangDesigningTheoryDrivenUserCentric2019a,ehsanHumanCenteredExplainableAI2020a}.

Fidelity is a popular metric, as most algorithmic research in XAI focuses on supervised machine learning (for a meta-survey, see~\cite{saeedExplainableAIXAI2023}).
In contrast, autonomous decision-making systems have received considerably less attention, as these systems are usually deployed in dynamically changing, often partially observable environments, where behavior may be difficult to explain even for humans. 
Although explainable reinforcement learning has worked to address the algorithmic challenges of explanation generation in this domain~\cite{milaniExplainableReinforcementLearning2024a}, the human-centered aspect has largely been unexplored.
As these systems often operate in social and safety-critical settings, understanding the cognitive processes of how humans explain autonomous behavior is essential for more effective XAI design and accurate trust calibration~\cite{dzindoletRoleTrustAutomation2003,mfaasCalibratingPedestriansTrust2021}.

\subsection{Framework of explanatory modes}\label{sec:tax}

Understanding the algorithms or cognitive processes of explanation is further complicated by the facts that different academic traditions have developed their own, sometimes conflicting nomenclature of causal reasoning and that the same words can have confusingly disparate folk meanings. 
To aid our discussion in this work and the broader context of XAI, we have attempted to unify many complex strands under the following framework of \textit{explanatory modes} (i.e., types of explanation).

\textbf{Teleological:} Makes reference to purpose, function, or an agent's goals, intentions, or desires, for example, following the pattern `x happened in order to bring about y'. This mode is also called intentional, purposive, or functional. We include purpose and function here because the `agent' may not be an explicit single agent but can be implied or distributed, as in phrases like `traffic laws are there to coordinate driver behavior and prevent accidents'.

\textbf{Mechanistic:} This mode follows the pattern `x happened because y happened' and implicitly or explicitly cites stable or abstract principles of how things work, assuming the same thing would happen again if the same conditions pertained. For example: the car stopped because it ran out of gas. This mode is also sometimes called \textit{causal}, but, as that word in folk usage means any reason, and so is the superclass of teleological and counterfactual, in this work we use \textit{mechanistic} any time we mean to emphasize factors which precede an effect.%

\textbf{Counterfactual:} An explanatory mode that references how events could have turned out differently from how they, in fact, did. For example, following the pattern `if x had been different, y would have happened'. In some literature, this mode is also called \textit{contrastive}; we use contrastive in our experiment instructions to avoid jargon.

\textbf{Descriptive:} Describes a situation by rephrasing what was observed, without reference to stable or abstract principles. For example: the car stopped because it ground to a halt. This mode is not strictly an explanation at all but is included here because people often offer as explanation what others would class as description, both in daily life and science~\cite{breiman2001statistical,craver2014towards,fuentes2024computational,miller2021breiman}.

\subsection{Why and how we apply this framework}

In this paper, we suggest that getting traction on how to give effective explanations to people about autonomous systems will involve a combination of integrating theoretical insights from cognitive science and conducting targeted empirical studies of how people generate and interpret explanations in context. 

Research in cognitive science reveals that generating and interpreting causal explanations involves sophisticated computations and inferences~\cite{lombrozo2006structure,byrne2007rational,byrne2023good,quillien2023counterfactuals,kirfel2022inference,navarre2024functional}.
In particular, humans often adopt an \textit{intentional stance}~\cite{dennett1987intentional} when they explain the behavior of a complex system, ascribing goals, beliefs, and intentions to the system, for example, `they went to the fridge because they wanted a beer [goal] and believed there was one in there [belief]'. These explanations are inherently teleological; they explain an agent's decision in terms of the purpose of that decision. 
In contrast, XAI systems usually generate mechanistic explanations that appeal to the mathematical, logical, or external processes and conditions involved in making a decision~\cite{speithReviewTaxonomiesExplainable2022}. 
An explanation of the inherent purpose of the decision is often lacking.

Furthermore, to understand how people interpret causal explanations, it is also important to assess whether people tend to give mechanistic or teleological explanations, even when the agent is not a person but a machine.
We also need to understand whether people's preferences for teleological and mechanistic explanations are at odds with applications of different theories of causation. 
This improved understanding would allow us to base the design of causal explanations on empirically validated principles. 

However, research in cognitive science has traditionally been more focused on tightly controlled environments, while XAI systems, especially in the case of autonomous decision-making, need to grapple with the complexities of real-world deployment. 
Therefore, rather than taking a simplified toy environment, we scope our experiments on decision-making for autonomous vehicles (AV), a popular and complex application domain for XAI~\cite{kuznietsovExplainableAISafe2024}.
Against this background, we aim to answer two research questions:

\begin{itemize}
    \item[\textbf{RQ1:}] Which modes of explanation do people prefer to use and receive when explaining behavior in the complex decision-making domain of autonomous driving?
    \item[\textbf{RQ2:}] How do people's preferences for explanatory modes change when the explained agent is an autonomous machine versus when it is a human?
\end{itemize}

To answer these questions, we discuss relevant research in cognitive science on causality, counterfactuals, and teleology as they relate to explanation and formulate hypotheses about the framework of explanatory modes.
To test these hypotheses, we designed an experiment consisting of two studies with human participants recruited through the online crowd-sourcing platform Prolific ($N_1=54$; $N_2=382$). 
In the first study, participants were asked to watch short driving scenarios with multiple interacting vehicles and then explain, in their own words, the behavior of a particular vehicle along the different explanatory modes.
In the second study, a different set of participants evaluated these explanations along various dimensions, such as perceived explanatory mode and perceived complexity, quality, and trustworthiness. 
This setup has the advantage of (i) generating explanations with a diverse sample of real people rather than writing them ourselves, (ii) having realistic situations where the ground truth is still relatively accessible to the explainer, and (iii) allowing us to explore various scenarios while keeping constant the overall context.

We find that the mode of explanation had a significant effect on the judgments of participants who preferred teleological and mechanistic concepts to counterfactual explanations.
They were also just as likely to refer to the mental states of AVs as human drivers.
In addition, perceived teleology was the best predictor of explanation quality and trustworthiness.
Based on these results, our primary recommendation to the field of XAI is to consider different explanatory modes as an important axis of analysis, especially focusing on the role and effect of teleology.
In summary, our main contributions are as follows:
\begin{itemize}
    \item Discussion of causality, counterfactuals, and teleology from cognitive science as they relate to explanation, highlighting the role of different \textbf{explanatory modes} (\cref{sec:foundations});
    \item Curation of a novel \textbf{dataset of human-generated and evaluated explanations} for autonomous driving, called the Human Explanations for Autonomous Driving Decisions (HEADD) dataset\footnote{The HEADD dataset and the code used for preprocessing and quantitative analyses are all available with documentation at \url{https://datashare.ed.ac.uk/handle/10283/8930}.} (\cref{sec:method});%
    \item A human participants experiment providing evidence that \textbf{teleology is preferred} by people when explaining agents' decision, regardless of whether the agent is perceived as human or machine (\cref{sec:quant-results});
    \item \textbf{Recommendations for the design and evaluation of XAI} in autonomous systems for better motivating and understanding the use of causality in human-centered XAI systems (\cref{sec:discussion}).
\end{itemize}

\section{Foundations of Explanation}\label{sec:foundations}

Causation is a cornerstone of effective explanation, and thus XAI.
Several fields investigate the notion of cause, albeit slightly differently: the XAI literature focuses on a distinction between causal explanations and counterfactuals, whereas the psychological literature has often made less of a difference and instead used counterfactuals as a means to study causal cognition, making an additional distinction between teleological explanations (focused on goals, intentions, or functions) versus causal-mechanistic ones~\cite{lombrozo2019mechanistic,joo2022understanding}.

\subsection{Causal and counterfactual explanations}\label{ssec:foundations:causal_explanations}

Explanation has a close relationship with causality~\cite{miller2019explanation} and, although there are nuances in the details of how each is formalized~\cite{halpern2016actual,hilton2007course,lewis2004causation}, it is broadly accepted that explaining a phenomenon usually involves asserting (some of) its cause(s). In turn, causality has a close relationship with counterfactuals~\cite{lewis2004causation,lagnado2013causal,quillien2023counterfactuals}. The counterfactual theory of causation, prominent in philosophy and psychology, holds that the meaning of `C caused E' is (roughly), that if C had not happened (but other factors had still played out the way they did) then E would not have happened~\cite{lewis2004causation,woodward2003making,pearl2000causality}. 

Although causal explanations implicitly involve counterfactual reasoning, it is nonetheless useful to distinguish a narrower meaning of causal explanations from counterfactual explanations. 
Counterfactual explanations explicitly highlight ways that things could have turned out differently (e.g., `If I had done \textit{x}, then \textit{y} would have happened'), whereas causal explanations as researchers often use the term refer to a chain of events (e.g., `\textit{y} happened because \textit{x} happened'). To avoid ambiguity, we here use the term \textit{mechanistic} for explanations that cite chains of events. 
Empirically, when people give mechanistic explanations they tend to focus on direct causes that co-vary with an outcome, for example, `a drunk driver caused the crash'~\cite{quillien2023counterfactuals}. 
When constructing counterfactuals, they tend to focus on controllable conditions that could have altered the outcome, for example `the crash would not have happened if the protagonist had driven home a different way'~\cite{mandel1996counterfactual}. 

Recent research has studied whether counterfactual or mechanistic explanations of an AI system are more effective. 
Empirical studies have found that users who are given a counterfactual explanation of a decision made by an autonomous system report more satisfaction with that explanation than users who are given a mechanistic explanation~\cite{celar2023people,warren2023categorical, warren2024explaining}. Counterfactual explanations are also more effective at improving the user's ability to predict the behavior of the system. For example, Celar and Byrne~\cite{celar2023people} showed participants the decisions made by an algorithm designed to determine whether someone's blood alcohol content (BAC) is above or below the legal limit for driving. The decisions were accompanied by a counterfactual explanation (`if the person had drunk 3 units of alcohol, they would be below the limit'), a mechanistic explanation (`drinking 5 units of alcohol caused the person to be above the limit'), or no explanation. Participants rated counterfactual explanations as more satisfying than mechanistic ones. This finding is intriguing when considering computational models of counterfactual reasoning which suggest people simulate several different counterfactual worlds as part of their process of generating explanations~\cite{lucas2015improved,quillien2022causal,quillien2023counterfactuals,droop2023extending}. 

However, the research that suggests an advantage for counterfactual explanations has focused on counterfactuals that typically highlight one possible alternative state of the world and on explanations generated by very simple algorithms. For example, determining whether someone is under the legal BAC limit can be done by applying simple rules. The question arises as to whether this advantage of counterfactuals generalizes to more complex settings. In particular, if the system is sufficiently complex, like a self-driving car, people might take an \emph{intentional stance} toward that system, conceiving of it as an agent. 

\subsection{Explanatory modes}\label{ssec:foundations:modes}

The human mind entertains different types of causal explanations~\cite{aristotle1933metaphysics,gerstenberg2017intuitive, joo2022understanding, byrne2023good}. 
Often, the same event can be expressed in different modes, particularly either in mechanistic terms (`the door opened because she kicked it') or in terms of the person's goals and desires (`she opened the door to let her friends in'). 
Explainable AI most often takes the former mechanistic stance, as the design of XAI methods is usually targeted at tracing the causal chain from input to output in terms of mathematical manipulations and algorithmic mechanisms.
In contrast, the latter example corresponds to taking an \emph{intentional stance}, whereby we can conveniently characterize and therefore predict an agent's behavior by attributing to them mental states such as beliefs, desires, and intentions~\cite{dennett1987intentional}. The intentional stance reliably emerges very early in development~\cite{gergely1995taking}, and its computational underpinnings are beginning to be mapped out by cognitive scientists~\cite{lucas2014child,baker2017rational,quillien2021simple}.

Explanations that use the intentional stance are teleological: they explain something in terms of the purpose it serves. For example, saying that Mary opened the fridge \emph{in order to} get some milk is a teleological explanation as it explains Mary's action in terms of its purpose. Teleological explanations are intuitive to the human mind, even outside the domain of psychological reasoning. They are readily produced and endorsed by children~\cite{kelemen1999rocks,lombrozo2016explanatory}. Adults sometimes endorse teleological explanations even for inanimate processes, for example, when under time pressure~\cite{kelemen2009human,kelemen2013professional}. Teleological explanations are generally useful because they identify causes that are \emph{robust} to changes in background circumstances: for example, my intention to drive home would have caused me to get home even if my usual route was closed, because I would then have taken a different route~\cite{lombrozo2006functional,lombrozo2006structure,lombrozo2019mechanistic,dennett1987intentional}. That is, under this conception the causes of an intentional system's behavior are its goals.

Evidence suggests that people can adopt the intentional stance toward artificial systems~\cite{perez2020adopting,clark2023social}. Whenever this is the case, teleological explanations of autonomous system decisions might be particularly effective, because they are consistent with the way the user intuitively represents the system~\cite{zerilli2022explaining, byrne2023good}.
In our study, we test whether this is the case for autonomous driving which involves explaining the coupled decision-making of multiple agents with a mixture of both human and artificial agents.
Will people prefer explanations of a self-driving car that are framed in teleological terms, or will they prefer more classic causal explanations (mechanistic or counterfactual)?
In the next section, we explain the relevance of teleology to the appropriateness of counterfactual explanations.

\subsection{Teleology and counterfactuals}

There is an interesting potential tension between counterfactual explanations and teleology. 
One recipe for generating counterfactual explanations is to take inspiration from the counterfactual theory of causation and produce a counterfactual of the form `if C had not happened, then E would not have happened' where C is the cause of outcome E. 
Consider, for example, a scenario where a self-driving car stops because pedestrians are crossing the road. 
A counterfactual explanation obeying the standard template would be `if the pedestrians were not crossing the road, the car would not have stopped'. 
This explanation effectively highlights the material cause of the car's behavior but does not have teleological content.

To generate a counterfactual explanation of the car's behavior in teleological terms, we might instead say `if the car had not stopped, it would have run over the pedestrians'. This counterfactual implicitly highlights the \emph{reason} for the car's behavior: the car stopped because if it had not, a bad consequence would have followed. Note that this kind of counterfactual has a different structure than the standard `if $\neg$C then $\neg$E' template: instead of altering the cause (the pedestrians crossing), we alter the effect (the car stopping). Teleological explanations are still implicitly causal: the car stopped because it computed that not stopping would have worse consequences than stopping. Nonetheless, the complexity of teleological explanations might mean that they will be difficult to express in terms of more standard counterfactuals. 

These considerations suggest the following prediction: if participants intuitively conceive of self-driving cars as agents, and apply the intentional stance toward them, they might not be satisfied by counterfactual explanations of their behavior, especially if these counterfactuals are of the form `if $\neg$C then $\neg$E'.

\subsection{The present study and hypotheses}\label{ssec:foundations:hypotheses}

We curate a dataset, called Human Explanations for Autonomous Driving Decisions (HEADD), of human-generated explanations of the decisions of AVs as well as evaluations of these explanations by a different set of participants. 
We anticipate that this dataset can shed light on a variety of questions regarding both explanation generation and interpretation. 
Below, we focus on our main predictions as they relate to the issues we reviewed in this section.

Our key experimental manipulation was the \emph{explanatory mode prompt}: the explanatory mode that participants were asked to use.
Our \textbf{Counterfactual}\footnote{We capitalize the explanatory mode when we refer to it as the independent variable of an experimental manipulation (i.e., prompt).} explanatory mode prompt requested participants to `describe changes to the scenario so that the blue vehicle takes different actions'. 
That is, it requested a counterfactual of the type `if $\neg$C then $\neg$E', which is difficult to interpret in teleological terms. 
Our \textbf{Mechanistic} explanatory mode prompt asked participants to `explain how the blue car was influenced in the scenario to take these actions'. 
This request elicits a causal explanation and does not specifically target teleological features of the situation (although it does not preclude them). 
Finally, our \textbf{Teleological} explanatory mode prompt requested participants to `explain why the blue car took these actions over different actions to reach its goal', foregrounding the agent's intentions, and emphasizing that the relevant counterfactuals are ones where the agent acts differently. 

Suppose there is a robust preference for counterfactual explanation for artificial systems.
In that case, we expect that explanations generated in response to the Counterfactual prompt should be rated as better than explanations generated in response to the other two prompts (replicating the results in~\cite{celar2023people,warren2023categorical}). 
In contrast, if the behavior of a complex agent (such as a self-driving car) activates an intentional stance, then the explanations generated in response to the Teleological (and possibly Mechanistic) prompt should be rated as better than the Counterfactual explanations. 
Additionally, if participants adopt an intentional stance, we predict that explanations containing more teleological features should be seen as more satisfying.
Therefore, our first two hypotheses test people's preferences for which explanatory modes they like to \emph{receive} when explaining complex autonomous behavior:

\begin{figure*}
    \centering
    \includegraphics[width=\linewidth]{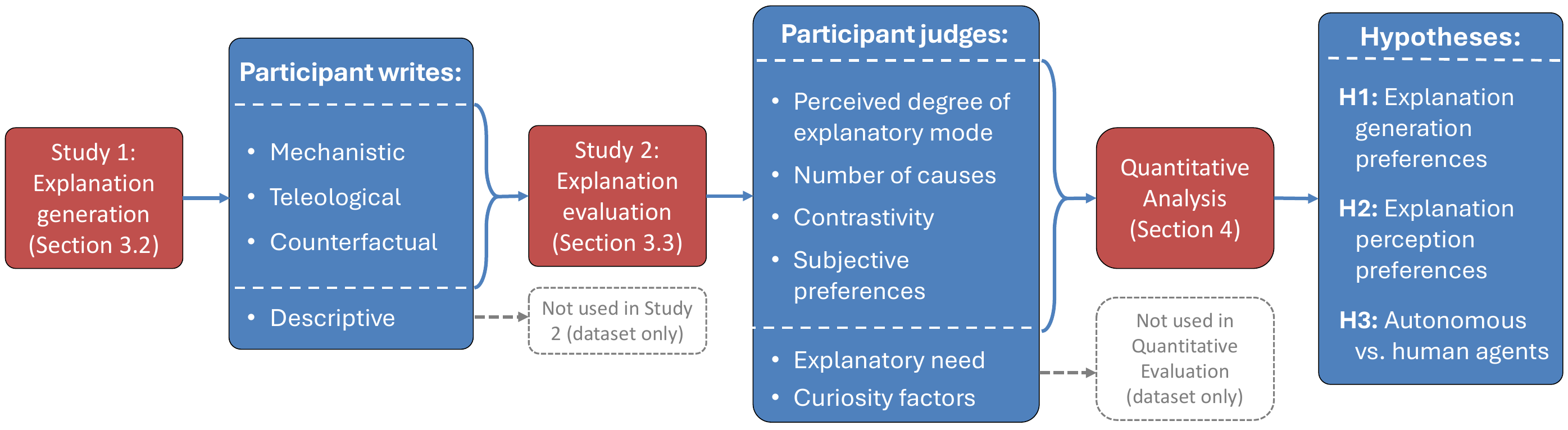}
    \caption{The workflow of our experiment with two studies. In the first study (\cref{ssec:method:1}), online participants generate explanations along the four explanatory modes. In the second study (\cref{ssec:method:2}), new participants evaluate the generated explanations along various subjective measures. Finally, we quantitatively analyze participant judgments and test our hypotheses (\cref{sec:quant-results}).}
    \label{fig:overall-workflow}
    \Description{Flowchart of the experiment. The figure consists of 6 stages. From left to right, these are: (1) Study 1: Explanation generation (Section 3.2); (2) Participant written explanation along the four explanatory modes (Mechanistic, Teleological, Counterfactual, Descriptive); (3) Study 2: Explanation evaluation (Section 3.2); (4) Participant judgments along the axes of perceived degree of explanatory mode, number of causes, contrastivity, subjective preferences, explanatory need, and curiosity factors; (5) Quantitative analysis (Section 4); (6) Hypotheses with H1 to H3.}
\end{figure*}

\begin{itemize}
    \item[\textbf{H1:}] Explanations given to the Teleological prompt are rated as better than the Counterfactual explanations. 

    \item[\textbf{H2:}] Explanations are rated as more satisfying if they are perceived to contain more teleological features.
\end{itemize}

Finally, we saw that people can readily form mental models of human agents by ascribing to them mental states such as beliefs, desires, and intentions~\cite{dennett1987intentional,gergely1995taking}. There is also evidence to suggest that people do this even for artificial systems~\cite{perez2020adopting,clark2023social}. 
Suppose that people do not ascribe mental states to artificial agents. In this case, we expect to see a difference between the effectiveness of teleological explanations when the explanations are generated about the perceived behavior of a human versus an autonomous system.
In contrast, if people do ascribe mental states to artificial systems, then we should not observe a significant difference between explanation ratings of human and autonomous agents.
Therefore, our third hypothesis tests people's preferences for teleological explanations with both human and autonomous agents:

\begin{itemize}
    \item[\textbf{H3:}] Given the same degree of perceived teleology, explanations of autonomous agents are rated as less satisfying than explanations of human agents.
\end{itemize}

\section{Experimental Methods}\label{sec:method}

Our experimental workflow is summarized in~\cref{fig:overall-workflow}.
The experiment consists of two human participants studies such that the first study provided the data used in the second study.
In the first study, participants generated natural language sentences about the behavior of a pre-selected vehicle in various driving scenarios along the four explanatory modes of~\cref{sec:tax}.
In the second study, we took these sentences and asked a different set of participants to evaluate the sentences as explanations according to their causal content and subjective quality.

For both studies, we designed surveys using the Qualtrics platform and relied on the online crowd-sourcing platform Prolific to recruit participants.\footnote{Qualtrics: \url{https://www.qualtrics.com/} --- Prolific: \url{https://www.prolific.com/} --- RoadRunner: https://www.mathworks.com/products/roadrunner.html}
We recruited from the USA and filtered for participants whose first language was English.
Participants were paid a pro-rated fee of £11 per hour and were shown an information sheet and our ethics approval.
Participants had to give their consent before they could access the studies. 
This study received ethics approval according to the Research Ethics Process of the School of Informatics, University of Edinburgh with reference \#282628.

Our design choices for both studies, including the phrasing of prompts and the number of iterations per study, were informed by both within-institution (i.e., internal) and crowdsourced (i.e., external) pilots.
For the first study, we received 10 internal participants and recruited 5 external participants.
For the second study, we received 13 internal participants and recruited 15 external participants for the pilots.

\subsection{Driving scenarios}

\begin{table*}
    \centering
    \caption{Summary table of driving scenarios. The columns Efficiency, Comfort, and Safety indicate to what degree the scenario was created with the intent to prompt an explanation related to the time efficiency, comfort, or safety of the driving action.}
    \label{tab:scenarios}
    \begin{tabular}{@{}l@{\quad}llll@{}}
    \toprule
         \textbf{\#} & \textbf{Summary} & \textbf{Efficiency} & \textbf{Comfort} & \textbf{Safety} \\
    \midrule
         1 & Cutting between cars on highway to exit through off-ramp. & Medium & High & High \\
         2 & Earlier right turn as oncoming vehicle stops to give way. & High & Low & Medium  \\
         3 & Merge into waiting line as oncoming vehicle leaves a gap. & High & Low & Medium \\
         4 & Enter two-lane roundabout early when oncoming vehicle enters outer lane. & High & Low & Medium  \\
         5 & Take over slow moving car likely looking for parking spot. & High & Low & Low \\
         6 & Overtake decelerating vehicle as it approaches T-junction. & High & Low & Medium \\
         7 & Cautious on-ramp to highway as bushes block view. & Low & Low & High \\
         8 & Carefully approaching crossroads with an occluded crossing. & Low & Low & High \\
         9 & Giving way in roundabout with occluded center until all clear. & Medium & Low & High \\
         10 & Slowly passing line of parked cars and stopping for one pulling out. & Low & Medium & High \\
         11 & Sudden breaking when a ball rolls onto the road from behind a truck. & Low & Medium & High \\
         12 & Passing between parked cars with truck ahead blocking the view. & Medium & Low & High \\
         13 & Slowing down for a high curvature turn. & Medium & High & Low \\
         14 & Rapid deceleration to merge behind truck on highway. & Low & High & Medium \\
         \bottomrule
    \end{tabular}
\end{table*}

\begin{figure*}
    \centering
    \includegraphics[width=0.31\textwidth]{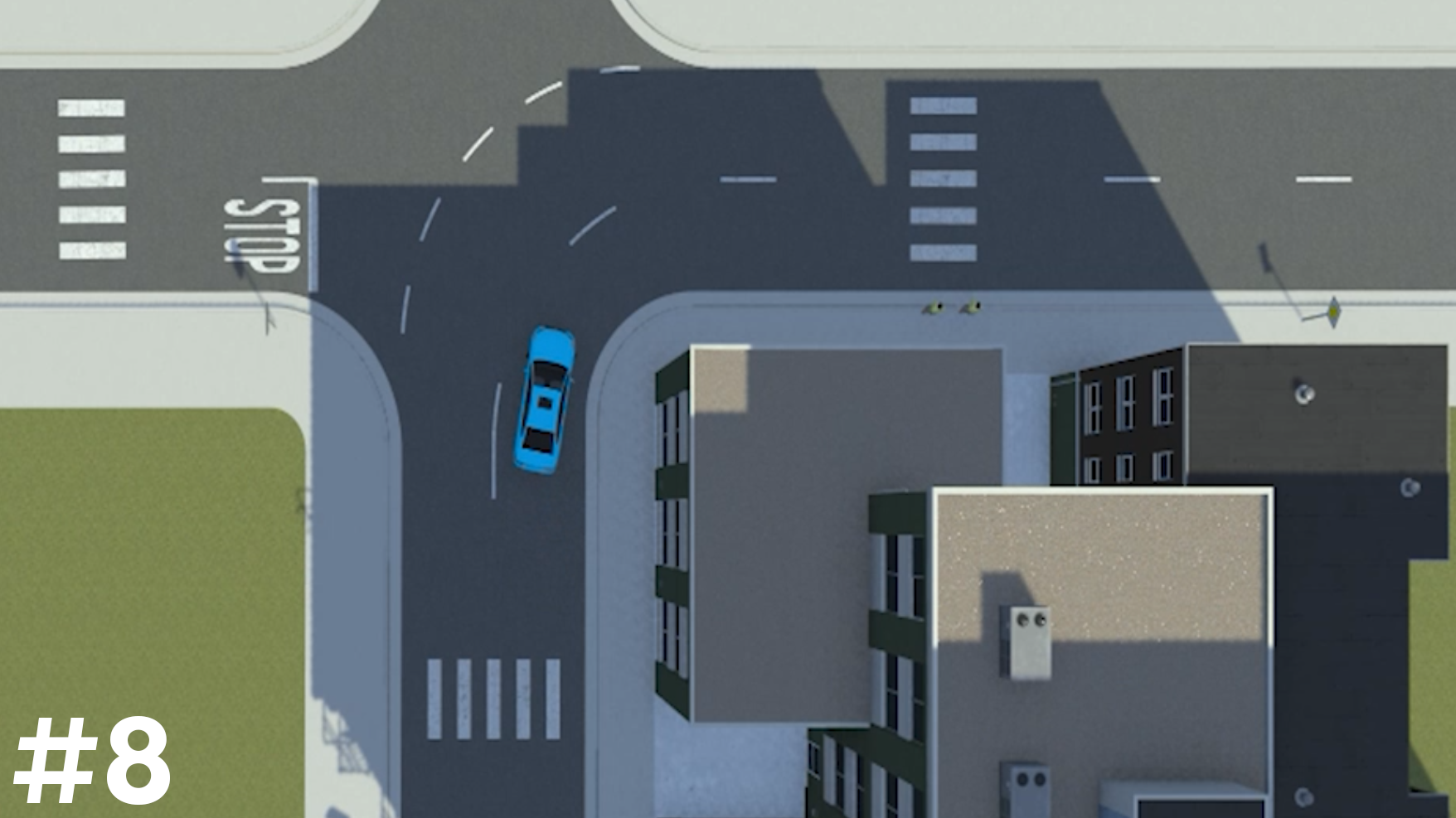}
    \hfill
    \includegraphics[width=0.31\textwidth]{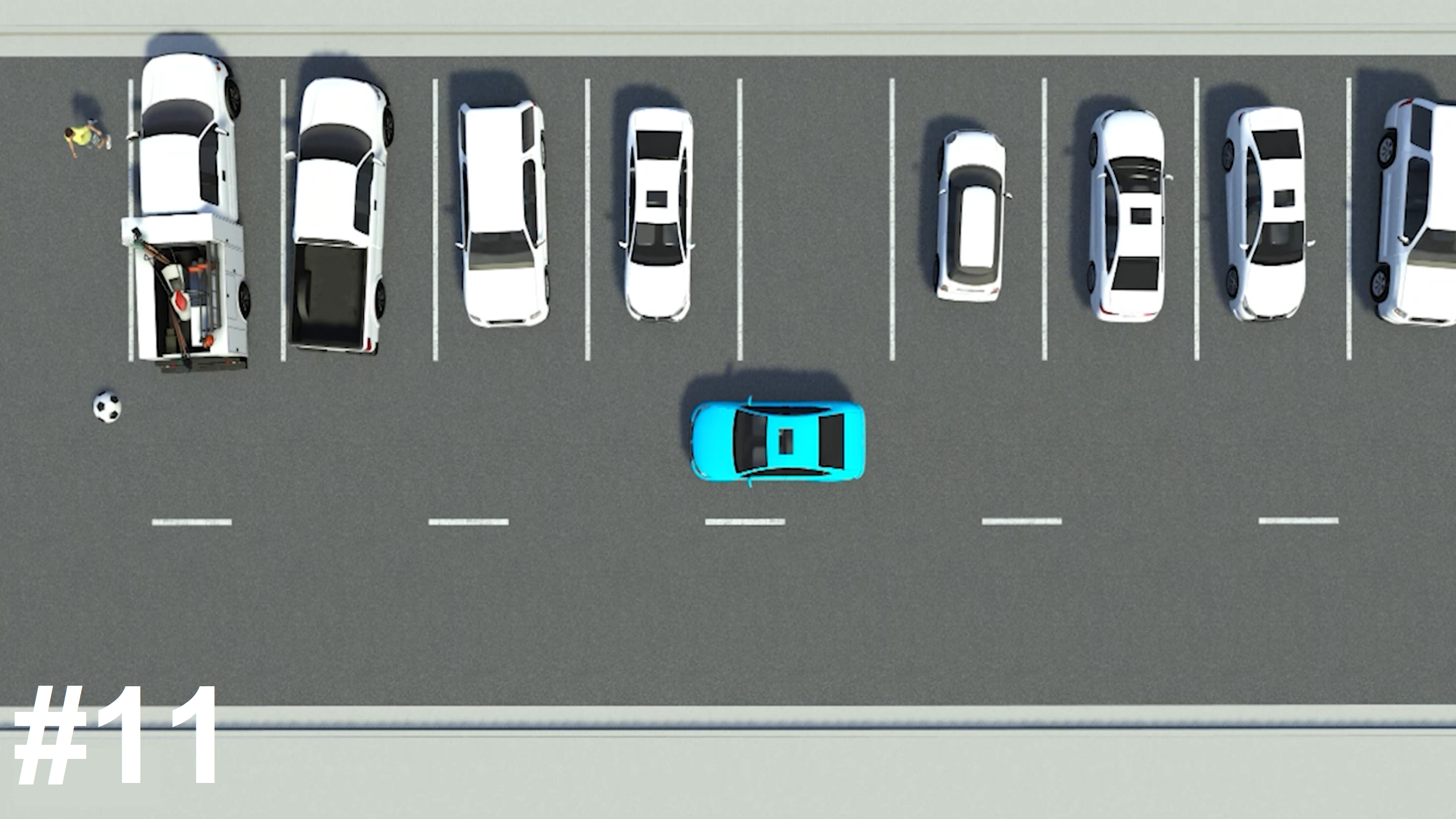}
    \hfill
    \includegraphics[width=0.31\textwidth]{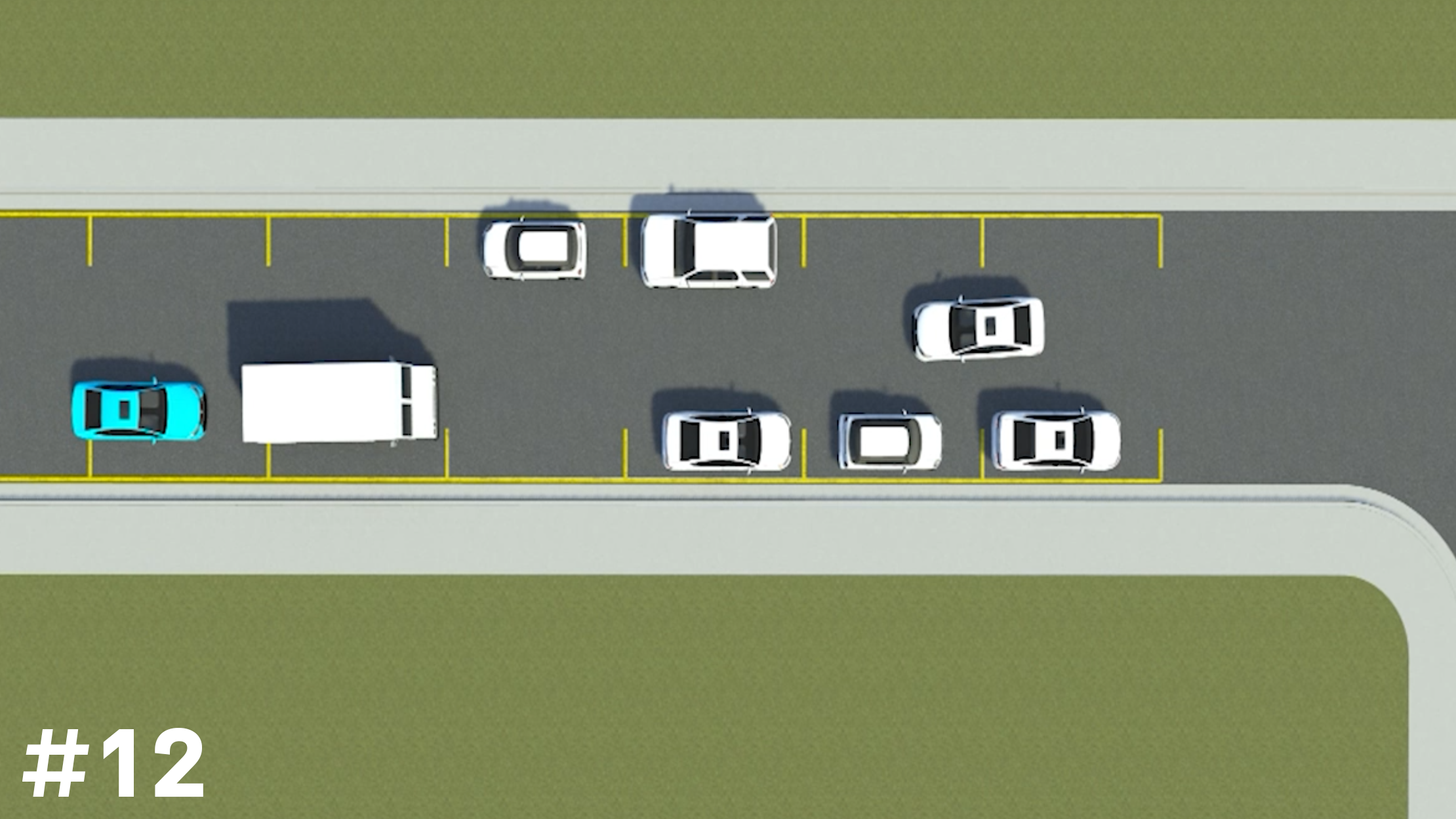}
    \caption{Three example scenarios. Participants were always asked to explain the behavior of the blue ego vehicle. (\textbf{Left}; \#8). The blue car slows down before turning right, as its view is blocked by a building. Once the view is clear, the blue car notices pedestrians at the crossing and stops. (\textbf{Mid}; \#11). The blue car is passing a row of parked cars when it perceives a ball rolling onto the road. It sharply breaks, as a child emerges from behind a truck. (\textbf{Right}; \#12) The blue car waits behind a truck obscuring its vision of the road. It keeps on waiting as the truck passes between the parked cars, to potentially avoid other oncoming cars.}
    \label{fig:example-scenarios}
    \Description{Three screenshots from videos of the driving scenarios #8, #11, and #12. Each image shows a bird's eye view snapshot of the blue autonomous ego vehicle driving in the scenario. The left image shows a curved main road with buildings blocking the view of a pedestrian crossing. The middle image shows a line of parked cars with the last car in the line blocking the view of a football and a child. The right image shows a narrow road with parked cars in each direction and a truck passing between the parked cars blocking the view of an oncoming vehicle.}
\end{figure*}

We designed 14 unique driving scenarios based on previous works from autonomous driving and XAI~\cite{albrechtInterpretableGoalbasedPrediction2021,hannaInterpretableGoalRecognition2021,wiegandExplanationThatExploring2020,kuznietsovExplainableAISafe2024} and used these 14 scenarios in both studies.
We picked scenarios to cover critical aspects of driving: safety, time efficiency, and comfort, to various degrees.
For example, some scenarios were designed to focus heavily on actions related to safety, while others were designed to focus more on reaching a goal faster, that is, time efficiency.
We summarize all 14 scenarios in~\cref{tab:scenarios} with full details available in the HEADD dataset.

We recreated each scenario from a top-down bird's eye view in the form of 5 to 15-second-long animated videos using the software RoadRunner 2023a by MathWorks.\footnotemark[3]
In each scenario, a single vehicle is highlighted in blue (hereafter, `ego vehicle').
Study 1 participants were always directed to explain the actions of the blue ego vehicle.
Example snapshots from such videos are shown in~\cref{fig:example-scenarios}.

\subsection{Study 1: Generating explanations}\label{ssec:method:1}

\subsubsection{Participants} 
A total of 54 participants (25 male, 27 female, 2 non-binary) filled out Study 1 with a median completion duration of approximately 27 minutes.
The participants' ages ranged between 19 to 73 years, with a median of 36 years. 
The majority of participants had some form of tertiary education (49 people) with the largest group having a Bachelor's degree (19 people).
Most participants reported having a valid driver's license (48 people) and the majority of participants had been driving for at least 2 years at the time of taking the experiment (44 people).

\subsubsection{Design}\label{sssec:method:1:design}

\begin{figure*}
    \centering
    \includegraphics[width=0.8\textwidth]{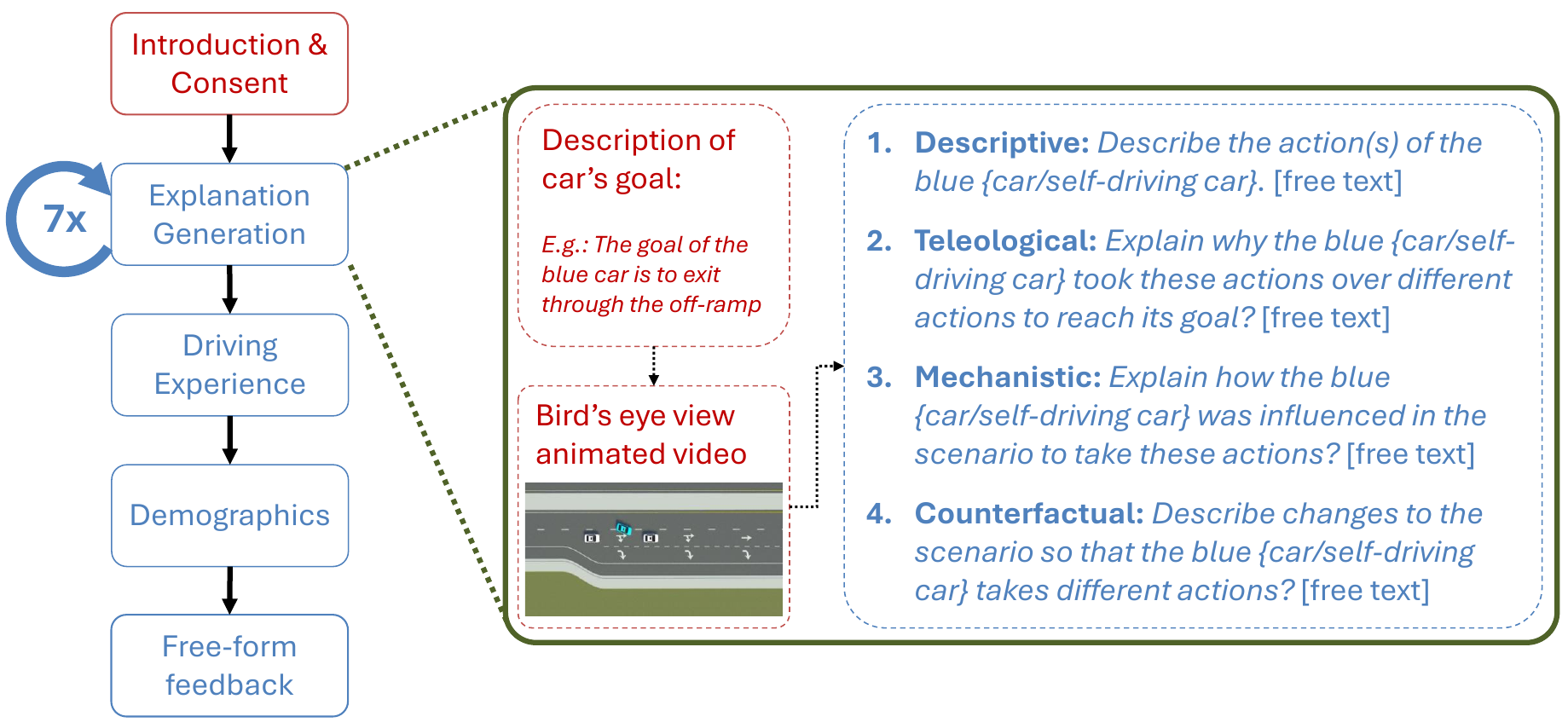}
    \caption{The workflow of Study 1. The text of each question (highlighted in italics) is copied verbatim as it appeared in the experiment to participants, with words in braces replaced according to the independent variables. Red boxes required participants to read and understand instructions. Blue boxes required input from the participants.}
    \label{fig:experiment-workflow-1}
    \Description{Flowchart of Study 1 in a top-to-bottom layout. On the left, there are five stages from start (top) to finish (bottom): (1) Introduction and Consent, (2) Explanation Generation, (3) Driving Experience, (4) Demographics, (5) Free-form Feedback. The explanation generation stage has an arrow pointing at itself with "7x" written in it to signal repetition 7 times. The explanation generation stage is expanded on the right side of the image showing the phrasing of the four explanatory mode prompts.}
\end{figure*}

The workflow of Study 1 is shown in~\cref{fig:experiment-workflow-1}.
Based on internal comments and the external completion rates in our pilots, we showed each participant seven driving scenarios, sampled randomly and evenly from the collection of 14 scenarios.
For each of the seven scenarios, participants were first shown the top-down animated video of the scenario and were told the goal (i.e., destination) of the blue vehicle.
Participants were then asked to answer four questions in their own words, corresponding to the four explanatory mode prompts as mentioned in~\cref{ssec:foundations:hypotheses} and further explained below under `Independent variables'.
Participants could re-watch the video and were allowed to answer the four questions in any order.\footnote{For analyses to rule out sequence effects, see~\cref{sssec:method:1:complexity}.}
We included two attention checks in the workflow and rejected 5 participants who failed at least one.
At the end of the experiment, participants were asked to answer questions regarding their driving experience (driving license, frequency, annual covered distance) and demographics (age, education level, gender).
We also gave them the opportunity to free form feedback.

\paragraph{Independent variables}\label{sssec:method:1:independent}
We used two independent between-subjects variables and one within-subjects variable to vary the experiment:
\begin{itemize}
    \item \textit{Scenario} [between-subjects]: Which seven scenarios were selected for the participant. Scenarios were sampled to enforce equal coverage for all 14 scenarios;
    \item \textit{AV} [between-subjects]: For half of all participants, we emphasized throughout the experiment that the ego vehicle was a `self-driving car' and for the other half it was simply a `car' in order to allow for measuring the effects of participants thinking that the ego vehicle was operated by an autonomous agent versus a human;
    \item \textit{Explanatory mode prompt} [within-subjects]: The requested explanatory mode for the explanation. We copy verbatim the text of each prompt:

    \begin{enumerate}
    \item \textit{Descriptive}: Describe the actions of the blue car;
    \item \textit{Teleological}: Explain why the blue car took these actions over different actions to reach its goal;
    \item \textit{Mechanistic}: Explain how the blue car was influenced in the scenario to take these actions;
    \item \textit{Counterfactual}: Describe changes to the scenario so that the blue vehicle takes different actions. (The new actions need not be the best actions in the scenario.)
\end{enumerate}

\end{itemize}

\paragraph{Dependent variables}\label{sssec:method:1:dependent}
Our dependent variable in Study 1 is the free-text response given to the four different explanatory mode prompts shown above.

\subsubsection{Data and preprocessing}
In total, we collected 1,447 sentences across all four explanatory modes. 
On finishing data collection, we first removed all participants who failed at least one attention check and then manually went through each sentence and removed meaningless or unusable responses (e.g., `I don't know', `None', `Nothing I can think of', etc.).
Over all scenarios, this process left 1,347 sentences and the following average number of sentences per explanatory mode per scenario (with standard deviation): Descriptive: $25.78 \pm 0.89$; Teleological: $25.00 \pm 1.24$; Mechanistic: $24.21 \pm 1.72$; and Counterfactual: $21.21 \pm 1.97$.
The average number of explanations given to the Counterfactual prompt is slightly lower than other Explanation prompts because in certain scenarios, especially \#13 and \#14, some participants struggled to think of a change that would alter the behavior of the ego vehicle.
Finally, we also corrected misspellings using a standard spell checker.

\subsubsection{Linguistic complexity}\label{sssec:method:1:complexity}
As we were also interested in investigating the correlation between linguistic complexity and the quality of explanations, we also performed the following linguistic preprocessing steps.
First, each explanation was processed using the Spacy library~\cite{spacy2}, which tokenized and lemmatized each sentence, and provided dependency parse trees.
An explanation may be composed of multiple sentences, in which case, the parsing was performed per sentence.
Second, we extracted standard measures of complexity for each explanation: the number of alphanumeric characters, tokens, unique lemmas, and sentences.
We calculated this both across the entire explanation and per sentence.
We also determined the average separation between dependent tokens in the dependency tree of a sentence as a measure of syntactic complexity.

\subsection{Study 2: Evaluating explanations}\label{ssec:method:2}

\subsubsection{Participants}
We recruited a total of 382 participants (187 male, 193 female, 2 non-binary) to participate in Study 2 with a median duration of completion of approximately 22 minutes.
The set of participants of Study 1 and 2 were mutually exclusive.
Participants' ages ranged between 19 to 83 years, with a median of 38 years. 
The majority of participants had some form of tertiary education (320 people) with the largest group having a Bachelor's degree (150 people).
Most participants reported having a valid driver's license (343 people) and the majority of participants had been driving for at least 2 years at the time of the experiment (353 people).

\subsubsection{Design}\label{sssec:method:2:design}

\begin{figure*}
    \centering
    \includegraphics[width=0.8\textwidth]{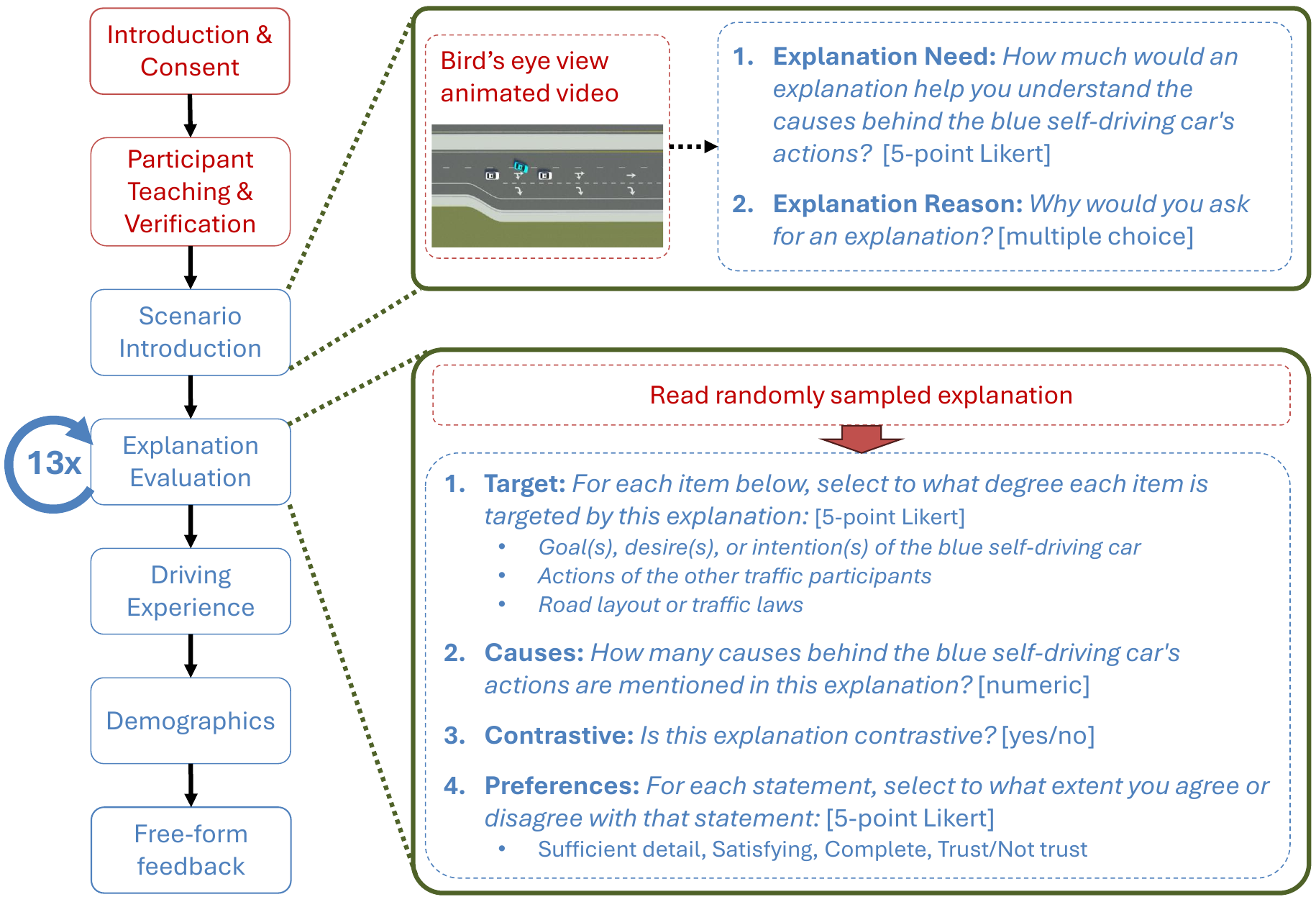}
    \caption{The workflow of Study 2. The text of each question (highlighted in italic) is verbatim as it appeared in the experiment. Red boxes required participants to read and understand instructions. Blue boxes required input from the participants.}
    \label{fig:experiment-workflow-2}
    \Description{Flowchart of Study 2 in a top-to-bottom layout. On the left, there are seven stages from start (top) to finish (bottom): (1) Introduction and Consent, (2) Participant Teaching and Verification, (3) Scenario Introduction, (4) Explanation Evaluation, (5) Driving Experience, (6) Demographics, (7) Free-form feedback. On the right side of the image, the stages Scenario Introduction and Explanation Evaluation are expanded to show the exact phrasing of prompts in the survey.}
\end{figure*}

The workflow of Study 2 is shown in~\cref{fig:experiment-workflow-2}.
Study 2 started with a teaching phase during which participants were guided through a simple driving scenario that explained to them the core concepts required to evaluate the explanations from Study 1.
This consisted of teaching them the various explanatory modes and a description of what causation is.
To keep Study 2 accessible, we aimed to minimize the amount of jargon used.
This means, for example, that we refer to explanatory modes as `explanation targets' and counterfactual as `contrastive'.
To verify whether participants had read and understood the content of the teaching phase, we asked them three single-choice questions regarding the definitions of explanatory mode, causation, and contrastivity. 
We rejected participants who could not answer any of the three questions correctly given two tries for each question.

Following the teaching phase, we picked at random one of the 14 scenarios, ensuring that all scenarios were picked evenly. 
We showed participants the video of the scenario without telling them the goal of the ego vehicle.
Participants were then asked to answer two questions in order for us to better understand the factors behind whether and what sort of explanations they wanted.

After this, we randomly sampled 13 explanations for the picked scenario.
We picked 13 based on feedback from the internal pilot and the completion rates in the external pilot.
We only included explanations from Study 1 which were given to the Teleological, Mechanistic, or Counterfactual prompt.
That is, Study 2 excluded the responses given to the Descriptive prompt of Study 1.
We did this because our focus is on understanding the causal content of explanations, but participants of Study 1 were not explicitly instructed to include causal information for the Descriptive prompt.
For each of the 13 explanations, participants were asked to answer several questions regarding its causal content and their subjective preferences, see `Dependent variables' below.

We included 3 attention checks and removed the data of all participants who failed at least one of them.
In total, we excluded 26 participants this way.
Finally, participants were asked the same driving experience and demographics-related questions as in Study 1 and had the opportunity to leave feedback.

\paragraph{Independent variables}
We manipulated two independent variables in Study 2. No other manipulations were performed as the goal of Study 2 was to provide rich evaluations of the explanations from Study 1:
\begin{itemize}
    \item \textit{Scenario} [between-subjects]: The scenario for which the participant had to evaluate the explanations;
    \item \textit{Explanation}: The 13 different non-descriptive explanations sampled randomly and evenly for the given scenario.
\end{itemize}

\paragraph{Dependent variables}\label{sssec:method:2:dependent}
We measured the following dependent variables in Study 2, named as they appear in our results in~\cref{sec:quant-results}. We copy verbatim the survey question text of each variable below.

First, we measured the need for an explanation for the given scenario. These were not used for our analyses later but are included in the dataset to enable more fine-grained future analyses:
\begin{enumerate}
    \item \textit{ExplanationNeed} [5-point Likert]: How much would an explanation help you better understand the causes behind the blue self-driving car's actions?
    \item \textit{ExplanationCuriosity} [multiple choice]: Why would you ask for an explanation? Select all that apply: (i) To know that I understand the self-driving car correctly; (ii) To understand what the self-driving car would do next; (iii) To know why the self-driving car did not make some other decision; (iv) To know what the self-driving car would have done if something had been different; (v) I was surprised by the self-driving car's actions and want to know what I missed.
\end{enumerate}

Second, we measured how much the displayed explanation targets an explanatory mode. We split the mechanistic explanatory mode into two scales to keep the question more accessible while also providing an opportunity for more fine-grained analysis:
\begin{enumerate}
    \setcounter{enumi}{2}
    \item \textit{TeleologyPerceived} [5-point Likert]: How much does the explanation target the goal(s), desire(s), or intention(s) of the blue car?
    \item \textit{MechanisticAgent} [5-point Likert]: How much does the explanation target the actions of the other traffic participants?
    \item \textit{MechanisticLayout} [5-point Likert]: How much does the explanation target the road layout or traffic laws?
    \item \textit{NumCauses} [non-negative integer]: How many causes behind the blue self-driving car's actions are mentioned in this explanation?
    \item \textit{Contrastive} [boolean]: Is this explanation contrastive?
\end{enumerate}

Third, we took preference judgments of the participants regarding the following aspects [all 5-point Likert]:
\begin{enumerate}
    \setcounter{enumi}{7}
    \item \textit{SufficientDetail}: This explanation of why the self-driving car behaved as it did has sufficient detail;
    \item \textit{Satisfying}: This explanation of why the self-driving car behaved as it did is satisfying;
    \item \textit{Complete}: This explanation of why the self-driving car behaved as it did seems complete;
    \item \textit{Trust}: This explanation lets me judge when I should trust and not trust the self-driving car.
\end{enumerate}

\subsubsection{Data and processing}

We collected 5,222 annotations for all 986 non-descriptive explanations generated in Study 1.
We filtered out the data of participants who failed at least one attention check but kept those who reported not having a valid driver's license as our results were not affected by their inclusion.
This gave 5 to 7 independent evaluations per explanation (Mean/Std: $5.3 \pm 0.58$).
To verify that we have sufficient statistical power, we ran a post-hoc power analysis for a $\chi^2$ goodness-of-fit test using the standard G*Power tool~\cite{faul2009statistical} with an effect size of 0.3, $\alpha$ error of 0.05, and 22 degrees of freedom: 14 for the independent variable Scenario and 8 for the Explanation (on average each scenario has 103 non-descriptive explanations from which we sample 13: $\lceil 103/13 \rceil=8$).
This gave a $1-\beta$ error probability of 0.966.

Using the linguistic data from the preprocessing of Study 1, we performed analyses to understand how linguistic complexity correlates with the perceived qualities of explanations.
We found that for each measure of complexity, our result did not change significantly depending on which measure we picked, therefore, in the following section we only report results using the number of tokens in the sentence.

\subsection{Statistical Analysis}
Data were analyzed in R v4.3.3 using RStudio version 2024.04.2+764. Package \textsl{lme4} \citep{bates2014fitting} was used to fit mixed-effect regression models following recommendations of \cite{meteyard2020best}, as is popular in behavioral sciences for performing multi-level modeling. Mixed-effect models allow one to quantify the \textit{main} or \textit{fixed} effects of interest (e.g., explanatory mode) while also accounting for variation in the effects across participants or stimuli. When reporting the results, the key metric is the slope $\beta$ of the relevant fixed effect and the corresponding confidence interval. We report the Akaike Information Criterion (AIC) which provides a relative measure of model fit (lower AIC indicates better fit) that penalizes more complex models.

\section{Results}\label{sec:quant-results}

\begin{table}
    \centering
    \caption{Example explanations with quality rankings (by Study 2 participants) for each explanation type, for scenario \#8 (see~\cref{fig:example-scenarios}). Quality ratings are shown in parentheses except for the descriptive mode as we did not collect ratings for those since the Descriptive prompt was not specifically designed to elicit causal reasoning (see~\cref{sssec:method:2:design}).}
    \renewcommand{\arraystretch}{1}
    \begin{tabular}{@{}p{0.2\linewidth}p{0.75\linewidth}@{}}
    \toprule
    \textbf{Explanation Type} & \textbf{Example Human-Generated Explanation}\\
    \midrule
    \textit{Descriptive} & `The blue car drove to the corner and turned right. It then stopped for two pedestrians at the crosswalk. Once the pedestrians were across it continued on its way.' \\
    {} & `The blue car took a right and waited at the crossing for pedestrians to cross before moving forward.' \\
    {} & `Turned right. Waited for a passenger to pass and then continued driving.' \\
    \midrule
    \textit{Teleological} & `The blue car took these actions to obey traffic laws and maintain the safety of the pedestrians crossing the street.' (4.8) \\
    {} & `This was the logical way to stay on the main road.' (3.33) \\
    {} & `It just followed the directions.' (2.06) \\
    \midrule
    \textit{Mechanistic} & `The blue car was influenced by the two pedestrians waiting to cross as it slowed to a complete stop allowing them to cross. The right turn was also 90 degrees which required the car to slow in order to make a successful turn.' (4.78) \\
    {} & `The markings on the road showed that the blue car needed to turn right in order to stay on the same road.' (3.4) \\
    {} & `It was influenced a bit by the zebra crossing.' (2) \\
    \midrule
    \textit{Counterfactual} & `If there were no pedestrians, then the car could have just immediately sped up to the speed limit instead of stopping in front of the crosswalk.' (4.33) \\
    {} & `The blue car could have increased its speed and not waited for the pedestrians but it could have resulted in an accident.' (3) \\
    {} & `The car could have just kept going maybe hitting the people walking.' (2.13)\\
    \bottomrule
    \end{tabular}
    \label{tab:scenarios-ex}
\end{table}

\Cref{tab:scenarios-ex} shows examples of participant-generated explanations for each explanatory mode prompt. As a manipulation check, we find that explanations generated to a Teleological prompt do exhibit more teleological features than explanations generated in response to other prompts, $p < .001$ for linear mixed model with random effects at the scenario, explanatory mode prompt, and participant levels. We also find that they are less likely to mention the actions of other agents, $p < .001$. In contrast, the type of explanatory mode prompt has no effect on the tendency of explanations to mention aspects of the road layout or traffic laws, $p = .44$.

\begin{figure}
    \centering
    \includegraphics[width=\linewidth]{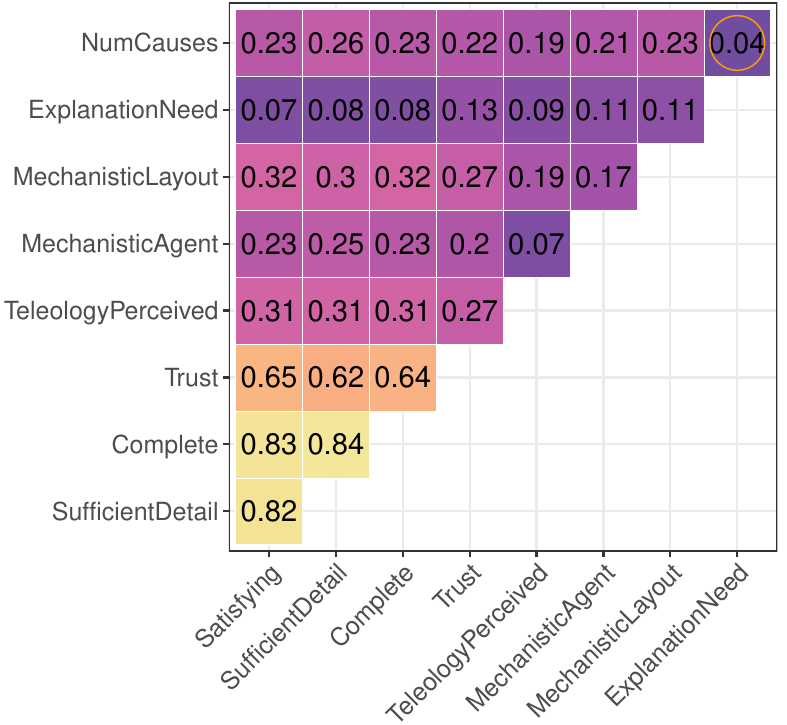}
    \caption{Zero-order correlation between ratings of Study 2. Correlation coefficients circled in orange are non-significant after Bonferroni correction. The Y-axis labels are the dependent variables of Study 2 (cf.,~\cref{sssec:method:2:dependent}); The most correlated variables are: (i) \textit{Satisfying}: whether the participants found the explanation satisfying; (ii) \textit{Complete}: perceived level of explanation completeness; (iii) \textit{SufficientDetail}: perceived level of detail in the explanation.} 
    \label{fig:correlationmatrix}
    \Description{Heatmap of an upper-triangular matrix of correlation coefficients between pairs of all dependent variables of Study 2. Each cell of the heatmap also shows the value of the correlation coefficient rounded to two decimals and a purple to yellow colormap is used to color each cell according to the magnitude of the coefficient.}
\end{figure}

Figure \ref{fig:correlationmatrix} displays the zero-order correlation matrix among judgments made by participants. Because ratings of Satisfyingness, Completeness, and Sufficient Detail were highly correlated with each other (all $r$ $>$ .8, all $p$ $<$ .001), we created a composite `Quality' variable by averaging them. This variable will be the main target in our analyses.

\subsection{Observations about the generated explanations of Study 1}\label{ssec:result:study-1}

Although a rigorous thematic analysis (e.g., coding) is outside the scope of this work, while initially exploring the explanations generated in Study 1, we made three qualitative observations which we present below.
We use scenario \#8 to illustrate the points here, however, the observations recur across all scenarios, which we confirm quantitatively.

\paragraph{Participants mix different explanatory modes in one explanation:}
While we wrote the prompts of Study 1 to target a single explanatory mode, many participants gave explanations that involved multiple modes. For example, in scenario \#8: `The blue car saw the crossing and slowed down after taking a right turn. It stopped for the pedestrians to cross.' Here, `seeing the crossing' is a mechanistic cause, while `stopping to let the pedestrians cross' is a teleological cause. To confirm this observation quantitatively, we filtered for all explanations according to participants' judgments in Study 2 along the dependent variables TeleologyPerceived, MechanisticAgent, MechanisticLayout, and Contrastive (see~\cref{sssec:method:2:dependent}). We used a minimum average ranking of 3.0 for the first three variables and majority voting for the last variable to determine whether the corresponding explanatory mode was present in the explanation. We found that 719 of the 986 (73\%) non-descriptive explanations contained at least two explanatory modes. %

\paragraph{Descriptive prompts often elicit teleological reasoning:} 
Many participants included teleological information when instructed purely to describe the behavior of the ego vehicle. We filtered the Descriptive explanations for the presence of purposive keywords\footnote{Including `wait to' (78), `allow to' (26), `to let' (26), `to exit' (17), `in order to' (13), `need to' (11), and `have to' (10), as well as their past tense conjugation.} and found 138 of the 361 explanations (38\%) in the Descriptive category contained references to the ego vehicle's intentions and goals. For example, in scenario \#8, one participant described the ego as follows: `The blue car took a right and waited at the crossing for pedestrians to cross before moving forward.' Here, `waiting for pedestrians to cross' is using the teleological explanatory mode. Descriptive explanations often include teleological (and only teleological) content, which highlights that people have a natural tendency to take an intentional stance even when they simply describe behavior.

\begin{figure}
    \centering
    \includegraphics[width=\linewidth]{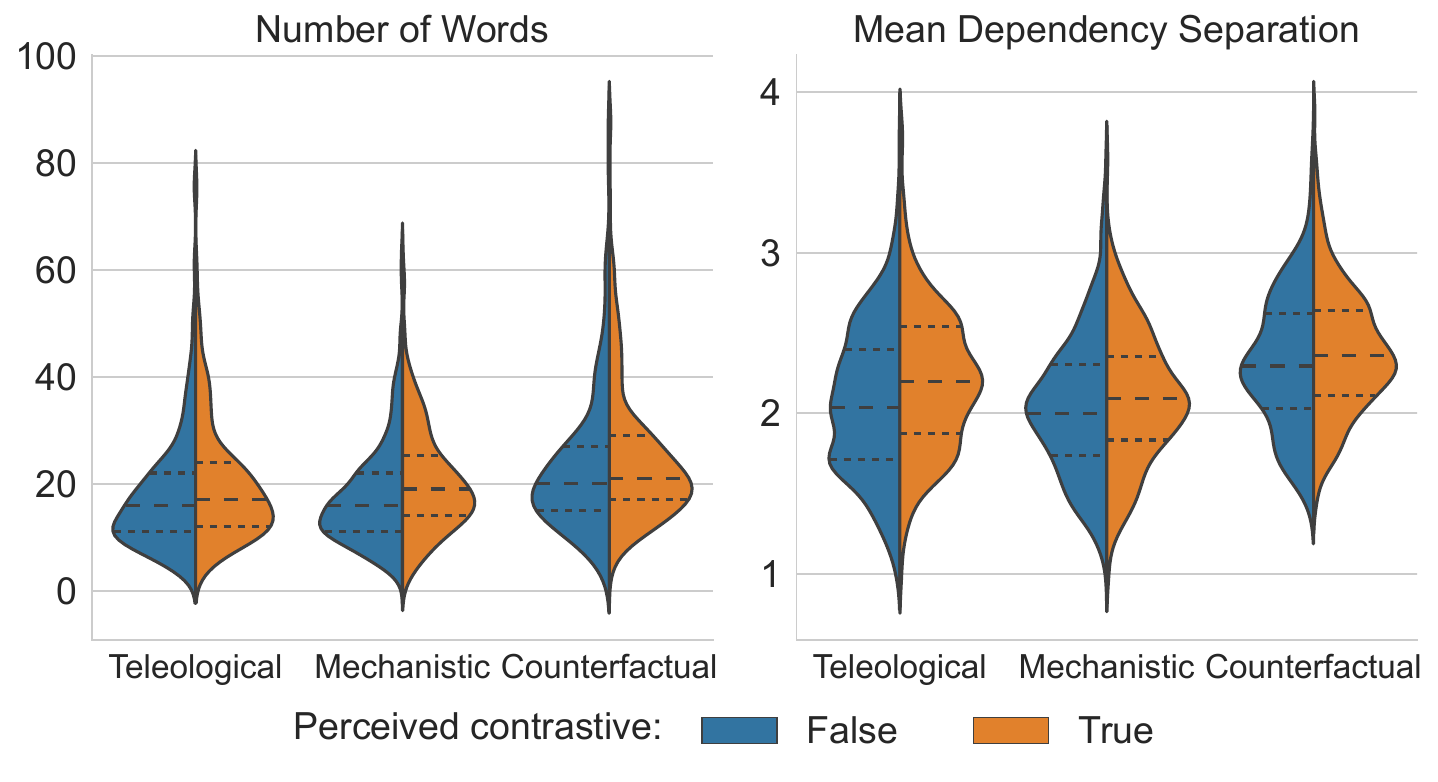}
    \caption{The distribution of the number of words and the mean dependency parse separation grouped by the explanatory mode prompts of Study 1 and colored by whether the explanation was perceived as contrastive by participants of Study 2. Lines inside the violin plots show the median and the interquartile ranges.}
    \label{fig:num-words-deplen}
    \Description{The figure consists of two subplots arranged horizontally. Both plots contain three violin plots. Each violin plot is split into two halves vertically based on whether an explanation was perceived as contrastive by participants of Study 2. The left plot shows the distribution of the number of words while the right plot shows the distribution of the mean separation of dependency relationships in the explanations given by participants of Study 1.}
\end{figure}

\paragraph{Counterfactual explanations tend to be more complex linguistically:}
\Cref{fig:num-words-deplen} shows the distribution of linguistic features of the responses generated in Study 1, as per the analysis in~\cref{sssec:method:1:complexity}.
While the average number of words is similar across the Teleological and Mechanistic explanatory modes, $p=.19, U=1589983$ in an independent Mann-Whitney U test, the Counterfactual prompt elicited longer sentences with a heavier tail, $p\ll.001, U=3833441$.
The average dependency separation is also higher for responses given to the Counterfactual prompt, $p \ll .001, U=4013531$, meaning that there tend to be more words between two dependent parts of the sentence. 
Additionally, explanations that were perceived as contrastive (i.e., given in response to the Counterfactual prompt) by participants of Study 2 have higher complexity measures, especially for the Teleological ($p<.001, U=440951$) and Mechanistic ($p<.001, U=394835.5$) prompt types.
One way to explain the higher linguistic complexity of counterfactual explanations is that they are typically expressed with conditionals, and this tends to result in long sentences (e.g. `If I had done X then Y would have happened').
Counterfactual reasoning is also argued to require sophisticated computations and more effort~\cite{byrne2007rational,quillien2023counterfactuals}, which may also result in more complex sentences.

\subsection{H1: Mechanistic and Teleological prompts lead to more satisfying explanations than Counterfactual prompts}\label{ssec:result:type}
\begin{figure}[t]
    \centering
    \includegraphics[width=\linewidth]{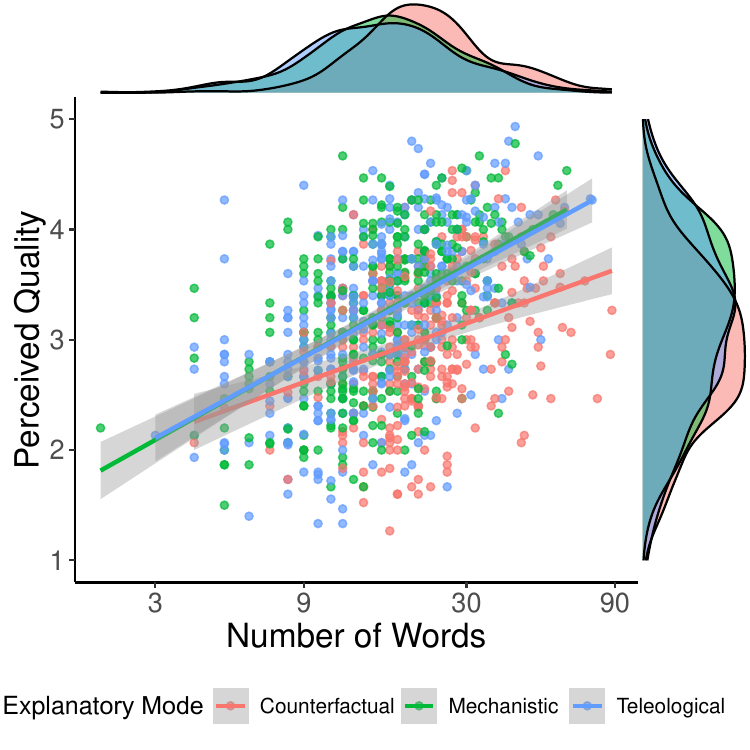}
    \caption{Perceived explanation quality as a function of the number of words in the explanation and the explanatory mode. Each dot corresponds to the average quality rating of one explanation (averaging the ratings of 5 to 7 participants). Lines are linear fits with shaded areas showing 95\% CIs.} 
    \label{fig:qualityByExp}
    \Description{Scatter plot of perceived quality rankings from Study 2 with fitted linear regression lines laid over the scatter plot for each explanatory mode prompt. The x-axis shows the number of words in an explanation, and the y-axis shows the corresponding Quality ranking. The estimated density function of the scatter plots for each explanatory mode prompt is shown above and on the right of the figure.}
\end{figure}

On average, explanations generated in response to a Mechanistic or Teleological prompt in Study 1 were perceived as better by participants in Study 2 than explanations generated for a Counterfactual prompt; see Figure \ref{fig:qualityByExp}. This effect was statistically significant, as assessed in a linear mixed model with random slopes at the scenario level, and random intercepts at the scenario, explanation, and participant levels: relative to Counterfactual explanations, both Mechanistic ($\beta = .10$, [95\% CI: .05, .14]) and Teleological prompts ($\beta=.09$, [95\% CI: .05, .13]) elicited higher Quality. 

To verify that this result is not due to participants putting in less effort when writing certain types of explanation, we addressed possible sequence effects by performing analyses in which we control for linguistic measures of complexity (see~\cref{sssec:method:1:complexity}), as a proxy for effort. We found that the effect of explanatory mode is robust to all these controls. We also inspected explanations and found that response quality was high in the experiment. 
We replicated this section's `satisfaction' analysis while including the linguistic complexity measure as a covariate to the linear mixed model reported above. We find that Mechanistic and Teleological explanations were still perceived as significantly better than Counterfactual explanations in every model, all 95\% CI lower-bounds > .04.

In contrast, there was only weak evidence for an effect of explanatory mode on Trust. Using a similar linear mixed modeling approach as above, we find that relative to Counterfactual explanations, Mechanistic ($\beta = .04$, [95\% CI: .01, .08]) and Teleological explanations ($\beta = .04$, [95\% CI: -.01, .09]) are perceived as only slightly more trustworthy.

All linguistic measures of complexity had a positive effect on perceived Quality, all $p$s $< .001$ for linear mixed models with random effects at the scenario, explanation prompt, and participant levels. Figure \ref{fig:qualityByExp} shows, for example, that longer explanations (as indexed by number of words) are rated as better. Interestingly, there was an interaction between number of words and explanatory mode: the number of words had a larger effect on perceived quality for Mechanistic and Teleological explanations relative to Counterfactual explanations, as shown by a linear mixed model with random intercepts at the scenario, explanation and participant levels, interaction effect: $p < .001$; while Counterfactual explanations tended to be more complex with on average the longest length (24.61 words; cf.,~\cref{ssec:result:study-1}). Taken together, these results suggest participants find counterfactual explanations less satisfying regardless of length. Intuitively, short explanations tend to be lower effort explanations that are perceived as worse regardless of the original explanation prompt, but participants who put extra effort generated better explanations in response to the Teleological and Mechanistic prompts. 

\begin{figure*}[t]
    \centering
    \includegraphics[width=\linewidth]{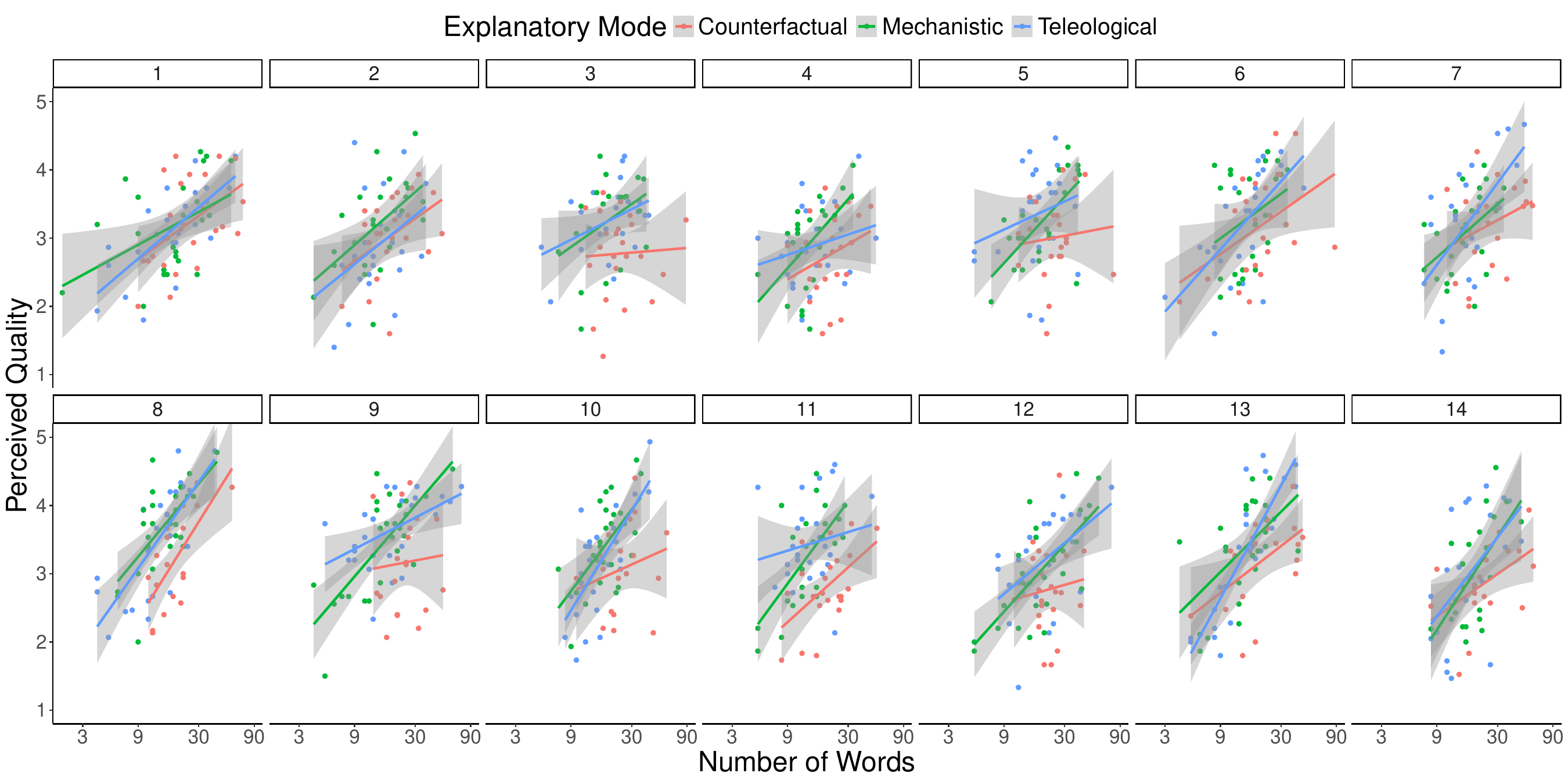}
    \caption{Perceived explanation quality as a function of number of words and explanatory mode, for each of the 14 scenarios. Each dot corresponds to the average quality rating of one explanation (computed by averaging the ratings of about 5 participants). Lines are linear fits with shaded areas showing 95\% CIs.} 
    \label{fig:qualityByExpScer}
    \Description{An array of fourteen plots each showing scatter plots and fitted linear regression lines for the fourteen scenarios for each explanatory mode prompt. Each plot in the array is identical in style to Figure 4 (without the density function estimates).}
\end{figure*}

Figure \ref{fig:qualityByExpScer} shows that the patterns discussed above are relatively robust across scenarios. Removing the scenario-level random slopes from the linear mixed model we used to test the effect of explanation type did not decrease model fit (full model, AIC = 15070, without random slopes, AIC = 15060). On the other hand, the effect of linguistic complexity appears to vary slightly depending on the scenario: removing the scenario-level random slopes from a linear mixed model predicting perceived Quality from number of words results in a slightly lower model fit (full model, AIC = 14900, without random slopes, AIC = 14908, $p = .003$).

\subsection{H2: Teleological features are the main predictor of perceived quality and trust}\label{ssec:result:perceived}
\begin{figure}
    \centering
    \includegraphics[width=\linewidth]{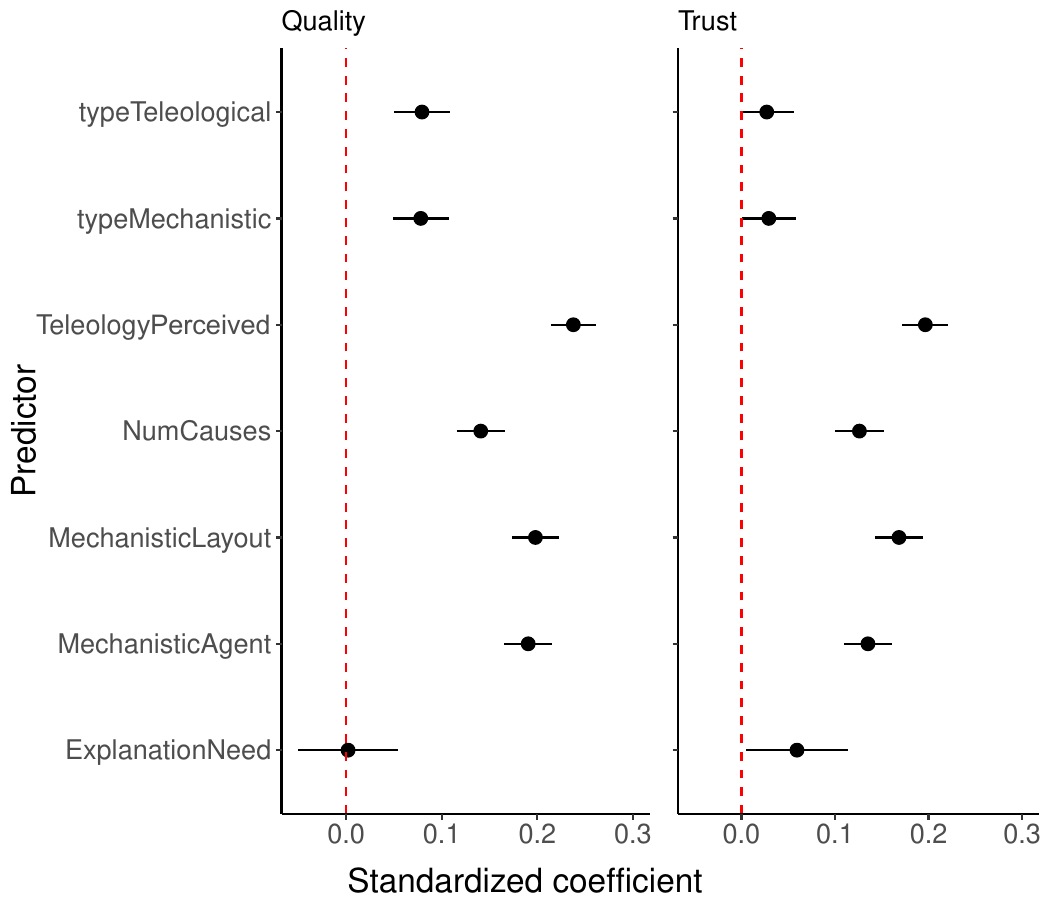}
    \caption{Standardized coefficients from linear mixed models predicting perceived Quality (left) and perceived Trust (right). The first two predictors (type) represent the experimental manipulation in Study 1 (the advantage of the Teleological and Mechanical prompts relative to the Counterfactual prompt, which is taken as a baseline; cf.,~\cref{sssec:method:1:dependent}), while the other predictors represent the effect of perceived features of explanations (cf.,~\cref{sssec:method:2:dependent}). Error bars represent 95\% CIs.} 
    \label{fig:predictorsCombined}
    \Description{Two side-by-side plots of regression coefficients showing mean and confidence intervals. The x-axis shows the value of the standardized coefficient. The y-axis shows the predictor variables. A dashed vertical red line marks zero on each plot.}
\end{figure}

Participants rated explanations along various feature dimensions: for example, whether an explanation mentioned the agent's goals, how many causes it described, etc. We ran linear mixed models to assess how well these features predicted participants' judgments of the Quality and Trust of explanations. \Cref{fig:predictorsCombined} shows the standardized coefficients from two linear mixed models, respectively predicting participants' judgments of Quality and Trust, with random intercepts at the scenario, explanation, and participant levels. 

Overall, perceived Teleology was the best predictor of perceived Quality and Trust: the explanations that participants judged as mentioning the goals, desires, or intentions of the agent were perceived as better and more trustworthy.
To more formally establish that Teleology is the best predictor of perceived Quality, we computed the AICs of linear mixed models where we removed either Teleology, MechanisticLayout, or MechanisticAgent as predictors. The model without Teleology had a substantially worse fit (AIC=14262) than the models without MechanisticLayout (AIC=14139) and without MechanisticAgent (AIC=14108). A similar approach yields the same result for perceived Trust (model without Teleology, AIC = 14712, without MechanisticLayout, AIC=14633, without MechanisticAgent, AIC=14577).
The number of causes mentioned, the extent to which the explanation mentioned the actions of other agents, and the extent to which it mentioned road layout or traffic laws, also reliably predicted both perceived Quality and Trust.

Importantly, perceived Teleology (how much an explanation mentioned the agent's desires, goals, and intentions, as judged by participants in Study 2) and the Teleology prompt (whether participants in Study 1 were explicitly instructed to write Teleological explanations) had independent effects on participants' Quality judgments: each variable has a significant effect when controlling for the other (see Figure \ref{fig:predictorsCombined}). Even for explanations generated in response to a Counterfactual or Mechanistic prompt, those that mentioned more teleological features were judged as better and more trustworthy. We did not find a difference in the effect of perceived Teleology across explanation types: adding random slopes at the explanation type level did not improve the fit of linear mixed models predicting perceived Quality ($p=.62$) or perceived Trust ($p=.27$). 

\subsection{H3: Neither perceived teleology nor quality ratings are affected by autonomous vs. human driver status}\label{ssec:result:automation}

\begin{figure}
    \centering
    \includegraphics[width=\linewidth]{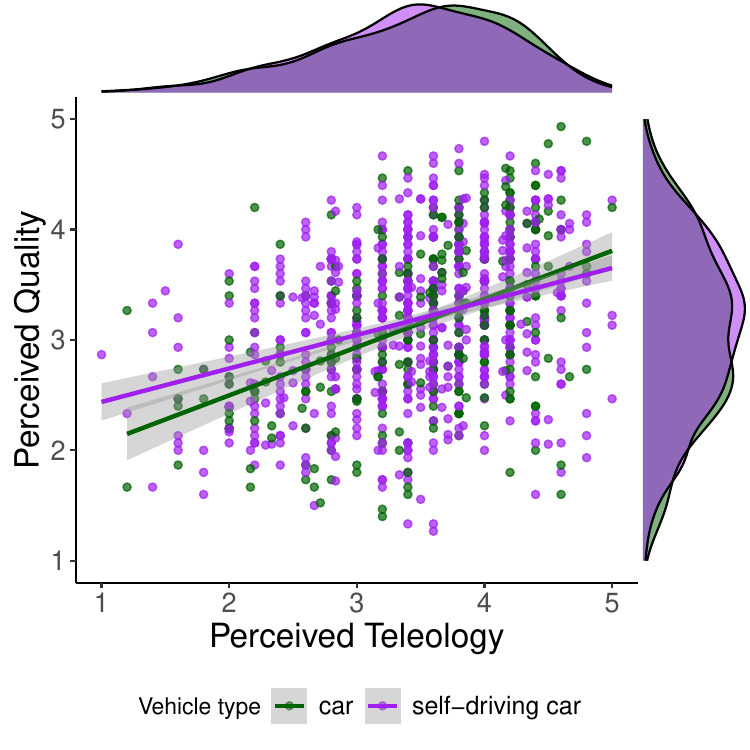}
    \caption{Perceived Quality as a function of Perceived Teleology, and whether the blue vehicle was identified as an autonomous vehicle. Each point represents one explanation. Lines are linear fits with 95\% confidence intervals.} 
    \label{fig:teleologyByHighlight}
    \Description{Scatter plot with two fitted regression lines laid over it. The x-axis shows the perceived teleology of explanations. The y-axis shows the perceived quality of the explanation. The fitted linear regression lines correspond to the AV manipulation and are largely overlapping.}
\end{figure}

We manipulated across conditions whether participants were told the blue car was an autonomous vehicle or was driven by a human driver (independent variable `AV' in Study 1). 
We then ran linear mixed models as above to assess whether the fact the person who generated the explanation in Study 1 was explaining the actions of a human or an autonomous vehicle had any effect on either Quality ratings or perceived Teleology (how much an explanation mentioned the agent's desires, goals and intentions, as judged by participants in Study 2). 
We found no significant difference in either ratings between the two conditions: including AV as a predictor variable contributed no significant improvement in the fit of the linear mixed models predicting perceived Quality ($p=.44$, $\chi^2$ test for goodness of fit) or perceived degree of Teleology ($p=.4$). Bayes Factors comparing the full linear mixed models to models omitting AV suggest evidence in favor of the null hypothesis (BF = .02 and BF = .02, respectively).
Furthermore, there was no improvement in model fit from including the interaction between AV and perceived Teleology ($p=.94$, BF = .09), indicating that the effect of perceived Teleology on perceived Quality is the same regardless of whether participants think the car is an autonomous vehicle; see Figure \ref{fig:teleologyByHighlight}.

\section{Discussion}\label{sec:discussion}

In this paper, we introduced a rich dataset of human evaluations of explanations, called the Human Explanations for Autonomous Driving Decisions (HEADD) dataset.
These explanations were human-generated and targeted the behavior of autonomous vehicles in short video clips.
We hope the HEADD dataset will be a valuable resource to help researchers understand how laypeople generate and interpret explanations.
Based on this dataset, our main result is that when people explain the behavior of self-driving vehicles, they often take an intentional stance, conceiving the vehicle as an agent with goals and beliefs. Specifically, we find that H1 and H2 are accepted and H3 is rejected:

\begin{enumerate}
    \item[\textbf{H1:}] Explanations to a Teleological prompt are judged as more satisfying than explanations generated in response to a Counterfactual prompt, which discourages teleological content;
    \item[\textbf{H2:}] Explanations that are perceived as having teleological content are judged as more satisfying---perceived teleology is the most important predictor of explanation satisfaction;
    \item[\textbf{H3:}] Whether people are explaining the behavior of human drivers or autonomous vehicles has no effect on the perceived quality of the explanations or the perceived teleology.
\end{enumerate}

The result for H3 suggests that people have no qualms about referring to autonomous vehicles as having beliefs, desires, and intentions. The intentional stance is not just leveled at people but can be a convenient abstraction to help us quickly conceptualize and refer to the outcome of any complex system that fulfills a function that could be expressed in a simplified way as `goal-like behavior'. This is evident in the way we talk, as witnessed by utterances like `the car doesn't want to start today', or `my laptop won't talk to the printer'. Even if people may not actually attribute mental content to machines, they still find it convenient to reason as if machines had mental states~\cite{clark2023social,perez2020adopting,zerilli2022explaining}.

Participants' preference for teleological explanation (H1, H2) highlights the usefulness of concepts from cognitive science for XAI~\cite{byrne2023good}. Cognitive scientists emphasize the fact that most of our knowledge is organized around domain-specific intuitive theories~\cite{hirschfeld1994mapping,gerstenberg2017intuitive}. 
Explanations that do not reflect the intuitive theory within which people understand a system might not be the most effective for conveying understanding. 
For autonomous driving, explanations that highlight reasons, objectives, and information states of the vehicles (i.e. that take the intentional stance) are likely to be more explanatory or satisfying to humans than those that focus on how the vehicles work.
Notably, when people adopt the intentional stance, they might not favor simple counterfactual-based explanations, in contrast to previous findings in simpler contexts~\cite{celar2023people,warren2023categorical}. 
A preference for the intentional stance also provides support for the design of decision-making systems that explicitly utilize a goal-oriented model (e.g.,~\cite{albrechtInterpretableGoalbasedPrediction2021,hannaInterpretableGoalRecognition2021}), as the decisions of these systems would be more amenable to explanation that follows human intuition.

In addition, our study contributes to a nascent literature in psychology that investigates `naturally occurring' explanations~\cite{zemla2017evaluating,sulik2023explanations}. While the psychological literature on explanations traditionally uses well-controlled stimuli, asking participants to evaluate a handful of experimenter-generated explanations, recent studies have asked participants to evaluate explanations collected from online forums~\cite{zemla2017evaluating} or collected from a crowd-sourcing platform~\cite{sulik2023explanations}. While these studies have focused on explanations that target general facts (e.g., `Why does thunder make noise?'), we contribute to this literature by exploring how people explain specific events (e.g., `Why did the car stop in this particular situation?'). More generally, our video stimuli depict scenes that are sufficiently rich to be interesting, but also simple enough that explainers can plausibly identify the reason for the agent's behavior. We also replicate some results from these previous studies, finding, for example, that more complex explanations (as indexed by the number of causes they mention) are more satisfying.

Taken together with existing research on human-centered XAI, our findings also have a number of implications for XAI in autonomous driving and, more broadly, for autonomous systems, which we discuss in the following subsections.

\subsection{Explanatory modes aid the purposeful design and evaluation of causality in XAI}\label{ssec:discussion:modes}
Our main recommendation is that the design and empirical evaluation of XAI that aims to be human-centered should consider explanatory modes as an important axis of analysis. Previous work has repeatedly shown that explanations that align with the cognitive processes of humans are also more easily interpreted~\cite{yangPsychologicalTheoryExplainability2022,kimHelpMeHelp2023,wangDesigningTheoryDrivenUserCentric2019a,miller2019explanation,millerExplainableAIDead2023,hirschfeld1994mapping,gerstenberg2017intuitive}. Still, there are many ways in which people give explanations: even a simple question like `Why does a pen have ink?' can be answered in multiple qualitatively different yet still causal ways (for example: `because someone filled the pen with ink', or `because the pen needs ink in order to perform its function')~\cite[Section 2.4]{miller2019explanation}. As we saw, these explanatory modes have a significant impact on the perceived quality and satisfyingness of the explanation. 

However, it is not only the evaluation of explanations that benefit from an increased understanding of the effects of explanatory modes but also their design. Most often the designers of XAI think along two axes: (i) the type of user query the explanation is supposed to answer~\cite{liaoQuestioningAIInforming2020}; and (ii) the algorithmic properties of the method that generates the explanation~\cite{schwalbe2023comprehensive}. While these axes are undoubtedly helpful in characterizing the explanation generation method, they are less effective in capturing the requirements on the causal content of the explanation. 

First, it is common practice in XAI to describe explanations as a function of user queries, capturing the causal content with various question-prototypes such as `what if'-questions, `why'-questions, or `why not'-questions~\cite{liaoQuestioningAIInforming2020}. A clear benefit of this approach is that humans can specify what interests them the most, however, as mentioned above, even a simple `why' question can be answered in multiple ways. Due to such ambiguities, explanatory modes could complement question prototypes to give both a human-centered and a more principled description of causal content. A simple first approach to test what combination works best would be to draw up a table with rows corresponding to question prototypes and columns to explanatory modes and create explanations for each table cell that fit the two properties. Designers can then select the best combination in a more structured way.

Second, design in XAI is often reduced to mere algorithm selection, such that properties like locality (`global' or `local') and model-specificity (`specific' or `agnostic') usurp all other considerations. This is made possible primarily by the prominence of out-of-the-box model interpretation methods, such as SHAP~\cite{shap}. Yet, blind reliance on these tools not only harms the human-centered usability of the system but also places bounds on the causal content that can be achieved. Take the example of SHAP, a local (i.e., input-dependent) and model-agnostic interpretation tool, which was built to generate mechanistic explanations. A recent survey found that people repeatedly apply SHAP in autonomous driving~\cite{kuznietsovExplainableAISafe2024} despite a range of mathematical~\cite{bilodeauImpossibilityTheorems2024,kumarProblemsShapleyvaluebasedExplanations2020} and usability-related~\cite{rudinStopExplainingBlack2019,kumarProblemsShapleyvaluebasedExplanations2020,ustunActionableRecourseLinear2019,poyiadziFACEFeasibleActionable2020} issues. Furthermore, there is no straightforward way to generate teleological explanations with it. Most papers never even consider whether mechanistic explanations are the most conducive to achieving a safer or more trustworthy driving experience. 

We believe that incorporating explanatory modes into the design process from the start could help mitigate the downstream effects of blindly relying on existing algorithms. The framework of explanatory modes can complement the already existing algorithmic taxonomies of XAI by providing a high-level perspective on the design of causal content. Should the explanation be purely descriptive? Should it focus on the goals of the system? Or should it highlight the logic involved in making a decision? While the answer will depend on the use case at hand, the framework of explanatory modes gives a good first-principles approach to beginning human-centered explanation design.

\subsection{Teleological explanations may be preferred} 
Our results indicate that for autonomous driving teleological explanations are most preferred by people.  This suggests that effective XAI systems that target autonomous systems might benefit from giving teleological explanations priority. One potential reason why people might prefer teleological explanations to other explanatory modes is that teleological explanations are robust to variation in the environment (e.g., if the goal of the agent is to reach destination X then it will try to achieve this regardless of whether a road en route is closed; see~\cref{ssec:foundations:modes}). Teleological explanations, therefore, have the benefit of better describing the outcomes of a complex system, which is essential to a better understanding, especially when dealing with safety-critical systems, such as autonomous driving.

More broadly, using teleological explanations is particularly relevant to the fields of explainable reinforcement learning~\cite{gyevnar2025objective,milaniExplainableReinforcementLearning2024a} and explainable AI planning~\cite{zhang2017plan} where explanations can easily be formulated in terms of the goals or terminal states of agents.
For example, existing work on model reconciliation in planning~\cite{sreedharanFoundationsExplanationsModel2021} and reward decomposition~\cite[e.g.,][]{septon2023integrating,juozapaitis2019explainable,lin2019distributional} directly consider goals and purpose and may benefit from framing their generated explanations within the explanatory modes framework, especially making clear the distinction between mechanistic (e.g., what minimal changes to a plan reconcile the differences with another plan) and teleological explanations (e.g., what changes to the goal state reconcile two different plans).

However, even for supervised ML, where there is no clearly specified terminal state or goal for the system, the analysis of explanatory modes can be relevant, because the system as a whole must have a purpose. For example, people may be well within their rights to expect explanations about why a particular credit application was refused, not just in terms of the feature attributions within an arbitrary logistic regression model (i.e., mechanistic explanation), but rather in terms of the monetary goals of the bank. Our framework of explanatory modes as given in~\cref{sec:tax} allows the analysis of explanations based on such higher-level goals because it is based on cognitive psychology and does not assume a particular computational methodology.

\subsection{Explanatory modes may be valuable for analyzing explainability pitfalls}
Based on the long-standing focus of XAI on mechanistic explanations, designers might think that people will prefer machines that are described in mechanistic terms (using non-agentic expressions such as `programming' and `directed'). However, our findings indicate that people like teleological explanations equally (using mentalistic expressions such as `intend to' and `want to') regardless of whether the agent was framed as a person or a self-driving car. This finding suggests that there is no penalty for using teleology to explain the behavior of artificial agents. 

However, there might be issues with a teleological framing. First, it may potentially be exploited as part of a dark design pattern~\cite{chromik2019dark}, wherein an actor frames the behavior of its agents in such a way that this downplays its responsibility in case of errors. For example, in a traffic accident, an explanation might portray the self-driving car's actions as if it had been taken by a human, using mentalistic expressions to try to direct attention away from its designers.

Second, while this manipulation may be done intentionally, we might accidentally manipulate users' perceptions through explainability pitfalls~\cite{ehsan2024explainability} (EPs). A teleological explanation in which the agent is framed as a human taking an intentional stance (e.g., `The car wanted to avoid the pedestrian') would likely contribute to increased anthropomorphization and bring with that a range of unexpected biases~\cite{duffy2003anthropomorphism}. Still, it is very difficult to predict how those biases could affect the downstream interactions and experiences of users. As Ehsan and Riedl~\cite{ehsan2024explainability} argue, the best designers can do is to set `a high bar of accountability [\dots] to be `pitfall aware' when designing XAI systems'. Unfortunately, as discussed in~\cref{ssec:discussion:modes}, XAI is often reduced to a blind reliance on interpretation tools, which directly ignores this recommendation. We believe that the analysis of explanatory modes is a valuable tool not only for expanding this myopic approach but also for charting the map of potential EPs in an XAI system, especially when it comes to making clear the differing effects of the mentalistic framing of agents in teleological explanations against the non-agentic framing of mechanistic ones.

\section{Limitations and Future Work}

The present study has some limitations. 
We chose to generate and evaluate explanations for the domain of autonomous driving. 
While this has the advantage of being accessible to a large demographic, it also partially limits the interpretation of our results to this domain. 
However, our insights from cognitive science are domain-independent, and future work should investigate both people's reliance on the intentional stance and the role of explanatory modes as an axis of analysis for better XAI.

Our experiment focuses on explanations of observed behavior rather than of a particular AI model, which brings with it the limitation that participants are not explaining in terms of the logic involved in the decision-making of the autonomous vehicle's actions.
Although, it is arguably impossible to explain in terms of the vast subsymbolic mathematical transformations of modern deep learning systems.
Still, it is unclear whether our `behavioral explanations' would be preferred if people knew more about the AI model driving the autonomous vehicle.
However, to create better explanations for people, one must still understand how they explain.
As such, the framework of explanatory modes and the supporting psychological experiment should still be very relevant to XAI. 

In~\cref{ssec:result:study-1}, we found that people tended to mix different explanatory modes in the same explanation. 
Given that we elicited explanations as free-text responses, it is impossible to have a clean separation between the various explanatory modes.
While this is an important consideration when designing an XAI system (i.e., should we mix explanatory modes in an automatically generated explanation?), to interpret our results, we do not need a clean separation.
Our findings show that teleology, regardless of the presence of other explanatory modes, tends to improve the perceived quality of the explanations (cf.,~\cref{fig:predictorsCombined}). 
A cleaner separation would likely make our claims about human preferences less robust because it would not be too similar to how people really explain behavior.

Furthermore, we only recruited participants from the US, which could have introduced a cultural bias.
As mentioned in~\cref{sec:intro}, different people interpret explanations differently based on individual context, so someone with a different cultural background to the US may prefer different explanatory modes. 
In addition, considering existing work which has shown that, at least on the level of ethics, cultural backgrounds play a significant role in how autonomous vehicles are perceived~\cite{bonnefonSocialDilemmaAutonomous2016,awadUniversalsVariationsMoral2020}, we believe it is important to investigate cultural effects on explanatory modes in the future.

Additionally, because driving is a generally well-understood domain, we expect the participants of our experiments to have pre-existing knowledge of the domain. Our findings should be interpreted in light of this context because in other less accessible autonomous decision-making domains, such as network routing or warehouse management, other explanatory modes might be preferred. We discuss the interplay between domain complexity and explanatory modes in~\cref{ssec:foundations:causal_explanations,ssec:foundations:modes}.

Following previous studies, we collected \emph{subjective} measures of explanation quality~\cite{zemla2017evaluating,sulik2023explanations,hoffmanMetricsExplainableAI2019}, such as how satisfying or trustworthy the listener considers the explanation. 
However, subjective measures can diverge from more objective measures (such as how much the explanation improves the listener's ability to predict the system) in subtle ways~\cite{bucinca2020proxy, van2021evaluating, kuhl2022keep, celar2023people, gyevnar2025objective}. 
Although we found teleological explanations were perceived as superficially more satisfying, it remains an open question to what extent they can help the user better predict the behavior of an autonomous vehicle, or better infer the details of what happened (cf.~\cite{kirfel2022inference,nam2023show, navarre2024functional}). 
Future research should also investigate the extent to which explanation preference varies across different contexts, beyond the range of scenarios considered here. 
It seems plausible that explanatory preferences might vary in function of many features of a situation, for example, whether an agent's goal is easy to infer, or how much the agent can see. 
Finally, despite our dichotomous framing, good natural explanations contain several elements of the modalities discussed, in various combinations.

\begin{acks}
The authors thank Nina Markl and the anonymous reviewers for their helpful feedback on previous drafts.
This work was supported in part by the UKRI Centre for Doctoral Training in Natural Language Processing, funded by the \grantsponsor{UKRI}{UK Research and Innovation}{https://www.ukri.org/} (grant \grantnum{UKRI}{EP/S022481/1}) and the University of Edinburgh, School of Informatics and School of Philosophy, Psychology \& Language Sciences.
\end{acks}

\bibliographystyle{ACM-Reference-Format}
\bibliography{refs,chi}


\begin{thebibliography}{101}


\ifx \showCODEN    \undefined \def \showCODEN     #1{\unskip}     \fi
\ifx \showISBNx    \undefined \def \showISBNx     #1{\unskip}     \fi
\ifx \showISBNxiii \undefined \def \showISBNxiii  #1{\unskip}     \fi
\ifx \showISSN     \undefined \def \showISSN      #1{\unskip}     \fi
\ifx \showLCCN     \undefined \def \showLCCN      #1{\unskip}     \fi
\ifx \shownote     \undefined \def \shownote      #1{#1}          \fi
\ifx \showarticletitle \undefined \def \showarticletitle #1{#1}   \fi
\ifx \showURL      \undefined \def \showURL       {\relax}        \fi
\providecommand\bibfield[2]{#2}
\providecommand\bibinfo[2]{#2}
\providecommand\natexlab[1]{#1}
\providecommand\showeprint[2][]{arXiv:#2}

\bibitem[Albrecht et~al\mbox{.}(2021)]%
        {albrechtInterpretableGoalbasedPrediction2021}
\bibfield{author}{\bibinfo{person}{Stefano~V. Albrecht}, \bibinfo{person}{Cillian Brewitt}, \bibinfo{person}{John Wilhelm}, \bibinfo{person}{Balint Gyevnar}, \bibinfo{person}{Francisco Eiras}, \bibinfo{person}{Mihai Dobre}, {and} \bibinfo{person}{Subramanian Ramamoorthy}.} \bibinfo{year}{2021}\natexlab{}.
\newblock \showarticletitle{Interpretable {{Goal-based Prediction}} and {{Planning}} for {{Autonomous Driving}}}. In \bibinfo{booktitle}{\emph{{{IEEE International Conference}} on {{Robotics}} and {{Automation}} ({{ICRA}})}}. \bibinfo{publisher}{IEEE Press}, \bibinfo{address}{Xi'an, China}, \bibinfo{pages}{1043–1049}.
\newblock
\href{https://doi.org/10.1109/ICRA48506.2021.9560849}{doi:\nolinkurl{10.1109/ICRA48506.2021.9560849}}


\bibitem[Aristotle(1933)]%
        {aristotle1933metaphysics}
\bibfield{author}{\bibinfo{person}{Aristotle}.} \bibinfo{year}{1933}\natexlab{}.
\newblock \bibinfo{booktitle}{\emph{Metaphysics}}. Vol.~\bibinfo{volume}{1}.
\newblock \bibinfo{publisher}{Harvard University Press}, \bibinfo{address}{Cambridge, MA}.
\newblock


\bibitem[Awad et~al\mbox{.}(2020)]%
        {awadUniversalsVariationsMoral2020}
\bibfield{author}{\bibinfo{person}{Edmond Awad}, \bibinfo{person}{Sohan Dsouza}, \bibinfo{person}{Azim Shariff}, \bibinfo{person}{Iyad Rahwan}, {and} \bibinfo{person}{Jean-Fran{\c c}ois Bonnefon}.} \bibinfo{year}{2020}\natexlab{}.
\newblock \showarticletitle{Universals and Variations in Moral Decisions Made in 42 Countries by 70,000 Participants}.
\newblock \bibinfo{journal}{\emph{Proceedings of the National Academy of Sciences}} \bibinfo{volume}{117}, \bibinfo{number}{5} (\bibinfo{date}{Feb.} \bibinfo{year}{2020}), \bibinfo{pages}{2332--2337}.
\newblock
\href{https://doi.org/10.1073/pnas.1911517117}{doi:\nolinkurl{10.1073/pnas.1911517117}}


\bibitem[Baker et~al\mbox{.}(2017)]%
        {baker2017rational}
\bibfield{author}{\bibinfo{person}{Chris~L Baker}, \bibinfo{person}{Julian Jara-Ettinger}, \bibinfo{person}{Rebecca Saxe}, {and} \bibinfo{person}{Joshua~B Tenenbaum}.} \bibinfo{year}{2017}\natexlab{}.
\newblock \showarticletitle{Rational quantitative attribution of beliefs, desires and percepts in human mentalizing}.
\newblock \bibinfo{journal}{\emph{Nature Human Behaviour}} \bibinfo{volume}{1}, \bibinfo{number}{4} (\bibinfo{year}{2017}), \bibinfo{pages}{0064}.
\newblock


\bibitem[Bates et~al\mbox{.}(2015)]%
        {bates2014fitting}
\bibfield{author}{\bibinfo{person}{Douglas Bates}, \bibinfo{person}{Martin Mächler}, \bibinfo{person}{Ben Bolker}, {and} \bibinfo{person}{Steve Walker}.} \bibinfo{year}{2015}\natexlab{}.
\newblock \showarticletitle{Fitting Linear Mixed-Effects Models Using lme4}.
\newblock \bibinfo{journal}{\emph{Journal of Statistical Software}} \bibinfo{volume}{67}, \bibinfo{number}{1} (\bibinfo{year}{2015}), \bibinfo{pages}{1--48}.
\newblock
\href{https://doi.org/10.18637/jss.v067.i01}{doi:\nolinkurl{10.18637/jss.v067.i01}}


\bibitem[Bilodeau et~al\mbox{.}(2024)]%
        {bilodeauImpossibilityTheorems2024}
\bibfield{author}{\bibinfo{person}{Blair Bilodeau}, \bibinfo{person}{Natasha Jaques}, \bibinfo{person}{Pang~Wei Koh}, {and} \bibinfo{person}{Been Kim}.} \bibinfo{year}{2024}\natexlab{}.
\newblock \showarticletitle{Impossibility theorems for feature attribution}.
\newblock \bibinfo{journal}{\emph{Proceedings of the National Academy of Sciences}} \bibinfo{volume}{121}, \bibinfo{number}{2} (\bibinfo{year}{2024}), \bibinfo{pages}{e2304406120}.
\newblock
\href{https://doi.org/10.1073/pnas.2304406120}{doi:\nolinkurl{10.1073/pnas.2304406120}}
\showeprint{https://www.pnas.org/doi/pdf/10.1073/pnas.2304406120}


\bibitem[Bonnefon et~al\mbox{.}(2016)]%
        {bonnefonSocialDilemmaAutonomous2016}
\bibfield{author}{\bibinfo{person}{Jean-Fran{\c c}ois Bonnefon}, \bibinfo{person}{Azim Shariff}, {and} \bibinfo{person}{Iyad Rahwan}.} \bibinfo{year}{2016}\natexlab{}.
\newblock \showarticletitle{The Social Dilemma of Autonomous Vehicles}.
\newblock \bibinfo{journal}{\emph{Science}} \bibinfo{volume}{352}, \bibinfo{number}{6293} (\bibinfo{date}{June} \bibinfo{year}{2016}), \bibinfo{pages}{1573--1576}.
\newblock
\href{https://doi.org/10.1126/science.aaf2654}{doi:\nolinkurl{10.1126/science.aaf2654}}


\bibitem[Breiman(2001)]%
        {breiman2001statistical}
\bibfield{author}{\bibinfo{person}{Leo Breiman}.} \bibinfo{year}{2001}\natexlab{}.
\newblock \showarticletitle{Statistical modeling: The two cultures (with comments and a rejoinder by the author)}.
\newblock \bibinfo{journal}{\emph{Statistical science}} \bibinfo{volume}{16}, \bibinfo{number}{3} (\bibinfo{year}{2001}), \bibinfo{pages}{199--231}.
\newblock


\bibitem[Bu\c{c}inca et~al\mbox{.}(2020)]%
        {bucinca2020proxy}
\bibfield{author}{\bibinfo{person}{Zana Bu\c{c}inca}, \bibinfo{person}{Phoebe Lin}, \bibinfo{person}{Krzysztof~Z. Gajos}, {and} \bibinfo{person}{Elena~L. Glassman}.} \bibinfo{year}{2020}\natexlab{}.
\newblock \showarticletitle{Proxy tasks and subjective measures can be misleading in evaluating explainable AI systems}. In \bibinfo{booktitle}{\emph{Proceedings of the 25th International Conference on Intelligent User Interfaces}} (Cagliari, Italy) \emph{(\bibinfo{series}{IUI '20})}. \bibinfo{publisher}{Association for Computing Machinery}, \bibinfo{address}{New York, NY, USA}, \bibinfo{pages}{454–464}.
\newblock
\showISBNx{9781450371186}
\href{https://doi.org/10.1145/3377325.3377498}{doi:\nolinkurl{10.1145/3377325.3377498}}


\bibitem[Byrne(2023)]%
        {byrne2023good}
\bibfield{author}{\bibinfo{person}{Ruth~M.J. Byrne}.} \bibinfo{year}{2023}\natexlab{}.
\newblock \showarticletitle{Good explanations in explainable artificial intelligence (XAI): evidence from human explanatory reasoning}. In \bibinfo{booktitle}{\emph{Proceedings of the Thirty-Second International Joint Conference on Artificial Intelligence}} \emph{(\bibinfo{series}{IJCAI '23})}. \bibinfo{publisher}{International Joint Conferences on Artifical Intelligence (IJCAI)}, \bibinfo{address}{Macao, P.R.China}, Article \bibinfo{articleno}{733}, \bibinfo{numpages}{9}~pages.
\newblock
\showISBNx{978-1-956792-03-4}
\href{https://doi.org/10.24963/ijcai.2023/733}{doi:\nolinkurl{10.24963/ijcai.2023/733}}


\bibitem[Byrne(2005)]%
        {byrne2007rational}
\bibfield{author}{\bibinfo{person}{Ruth M.~J. Byrne}.} \bibinfo{year}{2005}\natexlab{}.
\newblock \bibinfo{booktitle}{\emph{The Rational Imagination: How People Create Alternatives to Reality}}.
\newblock \bibinfo{publisher}{The MIT Press}, \bibinfo{address}{Cambridge, MA, USA}.
\newblock
\showISBNx{9780262269629}
\href{https://doi.org/10.7551/mitpress/5756.001.0001}{doi:\nolinkurl{10.7551/mitpress/5756.001.0001}}


\bibitem[Celar and Byrne(2023)]%
        {celar2023people}
\bibfield{author}{\bibinfo{person}{Lenart Celar} {and} \bibinfo{person}{Ruth~MJ Byrne}.} \bibinfo{year}{2023}\natexlab{}.
\newblock \showarticletitle{How people reason with counterfactual and causal explanations for Artificial Intelligence decisions in familiar and unfamiliar domains}.
\newblock \bibinfo{journal}{\emph{Memory \& Cognition}} \bibinfo{volume}{51}, \bibinfo{number}{7} (\bibinfo{year}{2023}), \bibinfo{pages}{1--16}.
\newblock
\href{https://doi.org/10.3758/s13421-023-01407-5}{doi:\nolinkurl{10.3758/s13421-023-01407-5}}


\bibitem[Chromik et~al\mbox{.}(2019)]%
        {chromik2019dark}
\bibfield{author}{\bibinfo{person}{Michael Chromik}, \bibinfo{person}{Malin Eiband}, \bibinfo{person}{Sarah~Theres V{\"o}lkel}, {and} \bibinfo{person}{Daniel Buschek}.} \bibinfo{year}{2019}\natexlab{}.
\newblock \showarticletitle{Dark Patterns of Explainability, Transparency, and User Control for Intelligent Systems}. In \bibinfo{booktitle}{\emph{IUI workshops}} \emph{(\bibinfo{series}{CEUR Workshop Proceedings}, Vol.~\bibinfo{volume}{2327})}. \bibinfo{publisher}{CEUR-WS.org}, \bibinfo{address}{Los Angeles, CA}, \bibinfo{numpages}{6}~pages.
\newblock


\bibitem[Clark and Fischer(2023)]%
        {clark2023social}
\bibfield{author}{\bibinfo{person}{Herbert~H Clark} {and} \bibinfo{person}{Kerstin Fischer}.} \bibinfo{year}{2023}\natexlab{}.
\newblock \showarticletitle{Social robots as depictions of social agents}.
\newblock \bibinfo{journal}{\emph{Behavioral and Brain Sciences}}  \bibinfo{volume}{46} (\bibinfo{year}{2023}), \bibinfo{pages}{e21}.
\newblock


\bibitem[Craver and Kaplan(2011)]%
        {craver2014towards}
\bibfield{author}{\bibinfo{person}{Carl~F. Craver} {and} \bibinfo{person}{David~M. Kaplan}.} \bibinfo{year}{2011}\natexlab{}.
\newblock \showarticletitle{Towards a Mechanistic Philosophy of Neuroscience}.
\newblock In \bibinfo{booktitle}{\emph{Continuum Companion to the Philosophy of Science}}, \bibfield{editor}{\bibinfo{person}{Steven French} {and} \bibinfo{person}{Juha Saatsi}} (Eds.). \bibinfo{publisher}{Continuum}, \bibinfo{address}{London, UK}, \bibinfo{pages}{268}.
\newblock


\bibitem[Dennett(1987)]%
        {dennett1987intentional}
\bibfield{author}{\bibinfo{person}{Daniel~C Dennett}.} \bibinfo{year}{1987}\natexlab{}.
\newblock \bibinfo{booktitle}{\emph{The intentional stance}}.
\newblock \bibinfo{publisher}{MIT Press}, \bibinfo{address}{Cambrdige, MA}.
\newblock


\bibitem[{Doshi-Velez} and Kim(2017)]%
        {doshi-velezRigorousScienceInterpretable2017}
\bibfield{author}{\bibinfo{person}{Finale {Doshi-Velez}} {and} \bibinfo{person}{Been Kim}.} \bibinfo{year}{2017}\natexlab{}.
\newblock \bibinfo{title}{Towards {{A Rigorous Science}} of {{Interpretable Machine Learning}}}.
\newblock \bibinfo{numpages}{13}~pages.
\newblock
\showeprint[arxiv]{1702.08608}~[cs, stat]


\bibitem[Droop and Bramley(2023)]%
        {droop2023extending}
\bibfield{author}{\bibinfo{person}{Stephanie Droop} {and} \bibinfo{person}{Neil~R Bramley}.} \bibinfo{year}{2023}\natexlab{}.
\newblock \bibinfo{title}{Extending counterfactual reasoning models to capture unconstrained social explanations}.
\newblock


\bibitem[Duffy(2003)]%
        {duffy2003anthropomorphism}
\bibfield{author}{\bibinfo{person}{Brian~R Duffy}.} \bibinfo{year}{2003}\natexlab{}.
\newblock \showarticletitle{Anthropomorphism and the social robot}.
\newblock \bibinfo{journal}{\emph{Robotics and autonomous systems}} \bibinfo{volume}{42}, \bibinfo{number}{3-4} (\bibinfo{year}{2003}), \bibinfo{pages}{177--190}.
\newblock


\bibitem[Dzindolet et~al\mbox{.}(2003)]%
        {dzindoletRoleTrustAutomation2003}
\bibfield{author}{\bibinfo{person}{Mary~T. Dzindolet}, \bibinfo{person}{Scott~A. Peterson}, \bibinfo{person}{Regina~A. Pomranky}, \bibinfo{person}{Linda~G. Pierce}, {and} \bibinfo{person}{Hall~P. Beck}.} \bibinfo{year}{2003}\natexlab{}.
\newblock \showarticletitle{The Role of Trust in Automation Reliance}.
\newblock \bibinfo{journal}{\emph{International Journal of Human-Computer Studies}} \bibinfo{volume}{58}, \bibinfo{number}{6} (\bibinfo{date}{June} \bibinfo{year}{2003}), \bibinfo{pages}{697--718}.
\newblock
\showISSN{1071-5819}
\href{https://doi.org/10.1016/S1071-5819(03)00038-7}{doi:\nolinkurl{10.1016/S1071-5819(03)00038-7}}


\bibitem[Ehsan et~al\mbox{.}(2024)]%
        {ehsanWhoXAIHow2024}
\bibfield{author}{\bibinfo{person}{Upol Ehsan}, \bibinfo{person}{Samir Passi}, \bibinfo{person}{Q.~Vera Liao}, \bibinfo{person}{Larry Chan}, \bibinfo{person}{I-Hsiang Lee}, \bibinfo{person}{Michael Muller}, {and} \bibinfo{person}{Mark~O Riedl}.} \bibinfo{year}{2024}\natexlab{}.
\newblock \showarticletitle{The {{Who}} in {{XAI}}: {{How AI Background Shapes Perceptions}} of {{AI Explanations}}}. In \bibinfo{booktitle}{\emph{Proceedings of the {{CHI Conference}} on {{Human Factors}} in {{Computing Systems}}}} \emph{(\bibinfo{series}{{{CHI}} '24})}. \bibinfo{publisher}{Association for Computing Machinery}, \bibinfo{address}{New York, NY, USA}, \bibinfo{pages}{1--32}.
\newblock
\showISBNx{9798400703300}
\href{https://doi.org/10.1145/3613904.3642474}{doi:\nolinkurl{10.1145/3613904.3642474}}


\bibitem[Ehsan and Riedl(2020)]%
        {ehsanHumanCenteredExplainableAI2020a}
\bibfield{author}{\bibinfo{person}{Upol Ehsan} {and} \bibinfo{person}{Mark~O. Riedl}.} \bibinfo{year}{2020}\natexlab{}.
\newblock \showarticletitle{Human-{{Centered Explainable AI}}: {{Towards}} a {{Reflective Sociotechnical Approach}}}. In \bibinfo{booktitle}{\emph{{{HCI International}} 2020 - {{Late Breaking Papers}}: {{Multimodality}} and {{Intelligence}}}} \emph{(\bibinfo{series}{Lecture {{Notes}} in {{Computer Science}}})}, \bibfield{editor}{\bibinfo{person}{Constantine Stephanidis}, \bibinfo{person}{Masaaki Kurosu}, \bibinfo{person}{Helmut Degen}, {and} \bibinfo{person}{Lauren {Reinerman-Jones}}} (Eds.). \bibinfo{publisher}{{Springer International Publishing}}, \bibinfo{address}{New York, NY}, \bibinfo{pages}{449--466}.
\newblock


\bibitem[Ehsan and Riedl(2024)]%
        {ehsan2024explainability}
\bibfield{author}{\bibinfo{person}{Upol Ehsan} {and} \bibinfo{person}{Mark~O. Riedl}.} \bibinfo{year}{2024}\natexlab{}.
\newblock \showarticletitle{Explainability pitfalls: Beyond dark patterns in explainable AI}.
\newblock \bibinfo{journal}{\emph{Patterns}} \bibinfo{volume}{5}, \bibinfo{number}{6} (\bibinfo{year}{2024}), \bibinfo{pages}{100971}.
\newblock
\showISSN{2666-3899}
\href{https://doi.org/10.1016/j.patter.2024.100971}{doi:\nolinkurl{10.1016/j.patter.2024.100971}}


\bibitem[Faul et~al\mbox{.}(2009)]%
        {faul2009statistical}
\bibfield{author}{\bibinfo{person}{Franz Faul}, \bibinfo{person}{Edgar Erdfelder}, \bibinfo{person}{Axel Buchner}, {and} \bibinfo{person}{Albert-Georg Lang}.} \bibinfo{year}{2009}\natexlab{}.
\newblock \showarticletitle{Statistical power analyses using G* Power 3.1: Tests for correlation and regression analyses}.
\newblock \bibinfo{journal}{\emph{Behavior research methods}} \bibinfo{volume}{41}, \bibinfo{number}{4} (\bibinfo{year}{2009}), \bibinfo{pages}{1149--1160}.
\newblock


\bibitem[Franklin et~al\mbox{.}(2021)]%
        {franklin2021blaming}
\bibfield{author}{\bibinfo{person}{Matija Franklin}, \bibinfo{person}{Edmond Awad}, {and} \bibinfo{person}{David Lagnado}.} \bibinfo{year}{2021}\natexlab{}.
\newblock \showarticletitle{Blaming automated vehicles in difficult situations}.
\newblock \bibinfo{journal}{\emph{iScience}} \bibinfo{volume}{24}, \bibinfo{number}{4} (\bibinfo{year}{2021}), \bibinfo{pages}{102252}.
\newblock
\showISSN{2589-0042}
\href{https://doi.org/10.1016/j.isci.2021.102252}{doi:\nolinkurl{10.1016/j.isci.2021.102252}}


\bibitem[Fuentes(2024)]%
        {fuentes2024computational}
\bibfield{author}{\bibinfo{person}{Jorge~Ignacio Fuentes}.} \bibinfo{year}{2024}\natexlab{}.
\newblock \showarticletitle{Computational systems as higher-order mechanisms}.
\newblock \bibinfo{journal}{\emph{Synthese}} \bibinfo{volume}{203}, \bibinfo{number}{2} (\bibinfo{year}{2024}), \bibinfo{pages}{1--26}.
\newblock


\bibitem[Gergely et~al\mbox{.}(1995)]%
        {gergely1995taking}
\bibfield{author}{\bibinfo{person}{Gy{\"o}rgy Gergely}, \bibinfo{person}{Zolt{\'a}n N{\'a}dasdy}, \bibinfo{person}{Gergely Csibra}, {and} \bibinfo{person}{Szilvia B{\'\i}r{\'o}}.} \bibinfo{year}{1995}\natexlab{}.
\newblock \showarticletitle{Taking the intentional stance at 12 months of age}.
\newblock \bibinfo{journal}{\emph{Cognition}} \bibinfo{volume}{56}, \bibinfo{number}{2} (\bibinfo{year}{1995}), \bibinfo{pages}{165--193}.
\newblock


\bibitem[Gerstenberg and Tenenbaum(2017)]%
        {gerstenberg2017intuitive}
\bibfield{author}{\bibinfo{person}{Tobias Gerstenberg} {and} \bibinfo{person}{Joshua~B Tenenbaum}.} \bibinfo{year}{2017}\natexlab{}.
\newblock \showarticletitle{Intuitive Theories}.
\newblock In \bibinfo{booktitle}{\emph{The Oxford Handbook of Causal Reasoning}}. \bibinfo{publisher}{Oxford University Press}, \bibinfo{address}{New York, NY, US}.
\newblock
\href{https://doi.org/10.1093/oxfordhb/9780199399550.001.0001}{doi:\nolinkurl{10.1093/oxfordhb/9780199399550.001.0001}}


\bibitem[Gyevnar et~al\mbox{.}(2023)]%
        {gyevnar2023transparencyGap}
\bibfield{author}{\bibinfo{person}{Balint Gyevnar}, \bibinfo{person}{Nick Ferguson}, {and} \bibinfo{person}{Burkhard Schafer}.} \bibinfo{year}{2023}\natexlab{}.
\newblock \showarticletitle{Bridging the Transparency Gap: What Can Explainable AI Learn From the AI Act?}. In \bibinfo{booktitle}{\emph{26th European Conference on Artificial Intelligence}}. \bibinfo{publisher}{IOS Press}, \bibinfo{address}{Krakow, Poland}, \bibinfo{pages}{964--971}.
\newblock


\bibitem[Gyevnar and Towers(2025)]%
        {gyevnar2025objective}
\bibfield{author}{\bibinfo{person}{Balint Gyevnar} {and} \bibinfo{person}{Mark Towers}.} \bibinfo{year}{2025}\natexlab{}.
\newblock \bibinfo{title}{Objective Metrics for Human-Subjects Evaluation in Explainable Reinforcement Learning}.
\newblock
\showeprint[arxiv]{2501.19256}~[cs.AI]
\urldef\tempurl%
\url{https://arxiv.org/abs/2501.19256}
\showURL{%
\tempurl}


\bibitem[Gyevnar et~al\mbox{.}(2024)]%
        {gyevnar2024cema}
\bibfield{author}{\bibinfo{person}{Balint Gyevnar}, \bibinfo{person}{Cheng Wang}, \bibinfo{person}{Christopher~G. Lucas}, \bibinfo{person}{Shay~B. Cohen}, {and} \bibinfo{person}{Stefano~V. Albrecht}.} \bibinfo{year}{2024}\natexlab{}.
\newblock \showarticletitle{Causal Explanations for Sequential Decision-Making in Multi-Agent Systems}. In \bibinfo{booktitle}{\emph{Proceedings of the 23rd International Conference on Autonomous Agents and Multiagent Systems}} (Auckland, New Zealand) \emph{(\bibinfo{series}{AAMAS '24})}. \bibinfo{publisher}{International Foundation for Autonomous Agents and Multiagent Systems}, \bibinfo{address}{Richland, SC}, \bibinfo{pages}{771–779}.
\newblock
\showISBNx{9798400704864}


\bibitem[Halpern(2016)]%
        {halpern2016actual}
\bibfield{author}{\bibinfo{person}{Joseph~Y Halpern}.} \bibinfo{year}{2016}\natexlab{}.
\newblock \bibinfo{booktitle}{\emph{Actual Causality}}.
\newblock \bibinfo{publisher}{MIT Press}, \bibinfo{address}{Cambridge, MA}.
\newblock


\bibitem[Hanna et~al\mbox{.}(2021)]%
        {hannaInterpretableGoalRecognition2021}
\bibfield{author}{\bibinfo{person}{Josiah~P. Hanna}, \bibinfo{person}{Arrasy Rahman}, \bibinfo{person}{Elliot Fosong}, \bibinfo{person}{Francisco Eiras}, \bibinfo{person}{Mihai Dobre}, \bibinfo{person}{John Redford}, \bibinfo{person}{Subramanian Ramamoorthy}, {and} \bibinfo{person}{Stefano~V. Albrecht}.} \bibinfo{year}{2021}\natexlab{}.
\newblock \showarticletitle{Interpretable Goal Recognition in the Presence of Occluded Factors for Autonomous Vehicles}. In \bibinfo{booktitle}{\emph{2021 IEEE/RSJ International Conference on Intelligent Robots and Systems (IROS)}} (Prague, Czech Republic). \bibinfo{publisher}{IEEE Press}, \bibinfo{address}{New York, NY, US}, \bibinfo{pages}{7044–7051}.
\newblock
\href{https://doi.org/10.1109/IROS51168.2021.9635903}{doi:\nolinkurl{10.1109/IROS51168.2021.9635903}}


\bibitem[Hilton and John(2007)]%
        {hilton2007course}
\bibfield{author}{\bibinfo{person}{Denis~J Hilton} {and} \bibinfo{person}{L~McClure John}.} \bibinfo{year}{2007}\natexlab{}.
\newblock \showarticletitle{The course of events: counterfactuals, causal sequences, and explanation}.
\newblock In \bibinfo{booktitle}{\emph{The psychology of counterfactual thinking}}. \bibinfo{publisher}{Routledge}, \bibinfo{address}{London, UK}, \bibinfo{pages}{56--72}.
\newblock


\bibitem[Hirschfeld and Gelman(1994)]%
        {hirschfeld1994mapping}
\bibfield{author}{\bibinfo{person}{Lawrence~A Hirschfeld} {and} \bibinfo{person}{Susan~A Gelman}.} \bibinfo{year}{1994}\natexlab{}.
\newblock \bibinfo{booktitle}{\emph{Mapping the mind: Domain specificity in cognition and culture}}.
\newblock \bibinfo{publisher}{Cambridge University Press}, \bibinfo{address}{Cambridge, UK}.
\newblock


\bibitem[Hoffman et~al\mbox{.}(2023)]%
        {hoffmanMetricsExplainableAI2019}
\bibfield{author}{\bibinfo{person}{Robert~R. Hoffman}, \bibinfo{person}{Shane~T. Mueller}, \bibinfo{person}{Gary Klein}, {and} \bibinfo{person}{Jordan Litman}.} \bibinfo{year}{2023}\natexlab{}.
\newblock \showarticletitle{Measures for explainable AI: Explanation goodness, user satisfaction, mental models, curiosity, trust, and human-AI performance}.
\newblock \bibinfo{journal}{\emph{Frontiers in Computer Science}}  \bibinfo{volume}{5} (\bibinfo{year}{2023}), \bibinfo{numpages}{15}~pages.
\newblock
\href{https://doi.org/10.3389/fcomp.2023.1096257}{doi:\nolinkurl{10.3389/fcomp.2023.1096257}}


\bibitem[Honnibal and Montani(2017)]%
        {spacy2}
\bibfield{author}{\bibinfo{person}{Matthew Honnibal} {and} \bibinfo{person}{Ines Montani}.} \bibinfo{year}{2017}\natexlab{}.
\newblock \bibinfo{title}{{spaCy 2}: Natural language understanding with {B}loom embeddings, convolutional neural networks and incremental parsing}.
\newblock
\urldef\tempurl%
\url{https://spacy.io/}
\showURL{%
\tempurl}


\bibitem[Joo et~al\mbox{.}(2022)]%
        {joo2022understanding}
\bibfield{author}{\bibinfo{person}{Sehrang Joo}, \bibinfo{person}{Sami~R Yousif}, {and} \bibinfo{person}{Frank~C Keil}.} \bibinfo{year}{2022}\natexlab{}.
\newblock \showarticletitle{Understanding “Why:” How implicit questions shape explanation preferences}.
\newblock \bibinfo{journal}{\emph{Cognitive Science}} \bibinfo{volume}{46}, \bibinfo{number}{2} (\bibinfo{year}{2022}), \bibinfo{pages}{e13091}.
\newblock


\bibitem[Juozapaitis et~al\mbox{.}(2019)]%
        {juozapaitis2019explainable}
\bibfield{author}{\bibinfo{person}{Zoe Juozapaitis}, \bibinfo{person}{Anurag Koul}, \bibinfo{person}{Alan Fern}, \bibinfo{person}{Martin Erwig}, {and} \bibinfo{person}{Finale Doshi-Velez}.} \bibinfo{year}{2019}\natexlab{}.
\newblock \bibinfo{title}{Explainable Reinforcement Learning via Reward Decomposition}.
\newblock


\bibitem[Keane et~al\mbox{.}(2021)]%
        {keane2021if}
\bibfield{author}{\bibinfo{person}{Mark~T. Keane}, \bibinfo{person}{Eoin~M. Kenny}, \bibinfo{person}{Eoin Delaney}, {and} \bibinfo{person}{Barry Smyth}.} \bibinfo{year}{2021}\natexlab{}.
\newblock \showarticletitle{If {{Only We Had Better Counterfactual Explanations}}: {{Five Key Deficits}} to {{Rectify}} in the {{Evaluation}} of {{Counterfactual XAI Techniques}}}. In \bibinfo{booktitle}{\emph{Proceedings of the {{Thirtieth International Joint Conference}} on {{Artificial Intelligence}}}}. \bibinfo{publisher}{{International Joint Conferences on Artificial Intelligence Organization}}, \bibinfo{address}{{Montreal, Canada}}, \bibinfo{pages}{4466--4474}.
\newblock


\bibitem[Kelemen(1999)]%
        {kelemen1999rocks}
\bibfield{author}{\bibinfo{person}{Deborah Kelemen}.} \bibinfo{year}{1999}\natexlab{}.
\newblock \showarticletitle{Why are rocks pointy? Children's preference for teleological explanations of th natural world.}
\newblock \bibinfo{journal}{\emph{Developmental psychology}} \bibinfo{volume}{35}, \bibinfo{number}{6} (\bibinfo{year}{1999}), \bibinfo{pages}{1440}.
\newblock


\bibitem[Kelemen and Rosset(2009)]%
        {kelemen2009human}
\bibfield{author}{\bibinfo{person}{Deborah Kelemen} {and} \bibinfo{person}{Evelyn Rosset}.} \bibinfo{year}{2009}\natexlab{}.
\newblock \showarticletitle{The human function compunction: Teleological explanation in adults}.
\newblock \bibinfo{journal}{\emph{Cognition}} \bibinfo{volume}{111}, \bibinfo{number}{1} (\bibinfo{year}{2009}), \bibinfo{pages}{138--143}.
\newblock


\bibitem[Kelemen et~al\mbox{.}(2013)]%
        {kelemen2013professional}
\bibfield{author}{\bibinfo{person}{Deborah Kelemen}, \bibinfo{person}{Joshua Rottman}, {and} \bibinfo{person}{Rebecca Seston}.} \bibinfo{year}{2013}\natexlab{}.
\newblock \showarticletitle{Professional physical scientists display tenacious teleological tendencies: purpose-based reasoning as a cognitive default.}
\newblock \bibinfo{journal}{\emph{Journal of experimental psychology: General}} \bibinfo{volume}{142}, \bibinfo{number}{4} (\bibinfo{year}{2013}), \bibinfo{pages}{1074}.
\newblock


\bibitem[Kim et~al\mbox{.}(2023)]%
        {kimHelpMeHelp2023}
\bibfield{author}{\bibinfo{person}{Sunnie S.~Y. Kim}, \bibinfo{person}{Elizabeth~Anne Watkins}, \bibinfo{person}{Olga Russakovsky}, \bibinfo{person}{Ruth Fong}, {and} \bibinfo{person}{Andr{\'e}s {Monroy-Hern{\'a}ndez}}.} \bibinfo{year}{2023}\natexlab{}.
\newblock \showarticletitle{"{{Help Me Help}} the {{AI}}": {{Understanding How Explainability Can Support Human-AI Interaction}}}. In \bibinfo{booktitle}{\emph{Proceedings of the 2023 {{CHI Conference}} on {{Human Factors}} in {{Computing Systems}}}} \emph{(\bibinfo{series}{{{CHI}} '23})}. \bibinfo{publisher}{Association for Computing Machinery}, \bibinfo{address}{New York, NY, USA}, \bibinfo{pages}{1--17}.
\newblock
\showISBNx{978-1-4503-9421-5}
\href{https://doi.org/10.1145/3544548.3581001}{doi:\nolinkurl{10.1145/3544548.3581001}}


\bibitem[Kirfel et~al\mbox{.}(2022)]%
        {kirfel2022inference}
\bibfield{author}{\bibinfo{person}{Lara Kirfel}, \bibinfo{person}{Thomas Icard}, {and} \bibinfo{person}{Tobias Gerstenberg}.} \bibinfo{year}{2022}\natexlab{}.
\newblock \showarticletitle{Inference from explanation.}
\newblock \bibinfo{journal}{\emph{Journal of Experimental Psychology: General}} \bibinfo{volume}{151}, \bibinfo{number}{7} (\bibinfo{year}{2022}), \bibinfo{pages}{1481}.
\newblock


\bibitem[Kuhl et~al\mbox{.}(2022)]%
        {kuhl2022keep}
\bibfield{author}{\bibinfo{person}{Ulrike Kuhl}, \bibinfo{person}{Andr\'{e} Artelt}, {and} \bibinfo{person}{Barbara Hammer}.} \bibinfo{year}{2022}\natexlab{}.
\newblock \showarticletitle{Keep Your Friends Close and Your Counterfactuals Closer: Improved Learning From Closest Rather Than Plausible Counterfactual Explanations in an Abstract Setting}. In \bibinfo{booktitle}{\emph{Proceedings of the 2022 ACM Conference on Fairness, Accountability, and Transparency}} (Seoul, Republic of Korea) \emph{(\bibinfo{series}{FAccT '22})}. \bibinfo{publisher}{Association for Computing Machinery}, \bibinfo{address}{New York, NY, USA}, \bibinfo{pages}{2125–2137}.
\newblock
\showISBNx{9781450393522}
\href{https://doi.org/10.1145/3531146.3534630}{doi:\nolinkurl{10.1145/3531146.3534630}}


\bibitem[Kumar et~al\mbox{.}(2020)]%
        {kumarProblemsShapleyvaluebasedExplanations2020}
\bibfield{author}{\bibinfo{person}{I.~Elizabeth Kumar}, \bibinfo{person}{Suresh Venkatasubramanian}, \bibinfo{person}{Carlos Scheidegger}, {and} \bibinfo{person}{Sorelle Friedler}.} \bibinfo{year}{2020}\natexlab{}.
\newblock \showarticletitle{Problems with {{Shapley-value-based}} Explanations as Feature Importance Measures}. In \bibinfo{booktitle}{\emph{Proceedings of the 37th {{International Conference}} on {{Machine Learning}}}}. \bibinfo{publisher}{PMLR}, \bibinfo{pages}{5491--5500}.
\newblock
\showISSN{2640-3498}


\bibitem[Kuznietsov et~al\mbox{.}(2024)]%
        {kuznietsovExplainableAISafe2024}
\bibfield{author}{\bibinfo{person}{Anton Kuznietsov}, \bibinfo{person}{Balint Gyevnar}, \bibinfo{person}{Cheng Wang}, \bibinfo{person}{Steven Peters}, {and} \bibinfo{person}{Stefano~V. Albrecht}.} \bibinfo{year}{2024}\natexlab{}.
\newblock \showarticletitle{Explainable AI for Safe and Trustworthy Autonomous Driving: A Systematic Review}.
\newblock \bibinfo{journal}{\emph{IEEE Transactions on Intelligent Transportation Systems}} \bibinfo{volume}{25}, \bibinfo{number}{12} (\bibinfo{date}{Oct.} \bibinfo{year}{2024}), \bibinfo{pages}{19342–19364}.
\newblock
\showISSN{1524-9050}
\href{https://doi.org/10.1109/TITS.2024.3474469}{doi:\nolinkurl{10.1109/TITS.2024.3474469}}


\bibitem[Lagnado et~al\mbox{.}(2013)]%
        {lagnado2013causal}
\bibfield{author}{\bibinfo{person}{David~A Lagnado}, \bibinfo{person}{Tobias Gerstenberg}, {and} \bibinfo{person}{Ro'i Zultan}.} \bibinfo{year}{2013}\natexlab{}.
\newblock \showarticletitle{Causal responsibility and counterfactuals}.
\newblock \bibinfo{journal}{\emph{Cognitive science}} \bibinfo{volume}{37}, \bibinfo{number}{6} (\bibinfo{year}{2013}), \bibinfo{pages}{1036--1073}.
\newblock


\bibitem[Lewis(2004)]%
        {lewis2004causation}
\bibfield{author}{\bibinfo{person}{David Lewis}.} \bibinfo{year}{2004}\natexlab{}.
\newblock \showarticletitle{Causation as influence}.
\newblock \bibinfo{journal}{\emph{The Journal of Philosophy}} \bibinfo{volume}{97}, \bibinfo{number}{4} (\bibinfo{year}{2004}), \bibinfo{pages}{182--197}.
\newblock


\bibitem[Liao et~al\mbox{.}(2020)]%
        {liaoQuestioningAIInforming2020}
\bibfield{author}{\bibinfo{person}{Q.~Vera Liao}, \bibinfo{person}{Daniel Gruen}, {and} \bibinfo{person}{Sarah Miller}.} \bibinfo{year}{2020}\natexlab{}.
\newblock \showarticletitle{Questioning the {{AI}}: {{Informing Design Practices}} for {{Explainable AI User Experiences}}}. In \bibinfo{booktitle}{\emph{Proceedings of the 2020 {{CHI Conference}} on {{Human Factors}} in {{Computing Systems}}}} \emph{(\bibinfo{series}{{{CHI}} '20})}. \bibinfo{publisher}{Association for Computing Machinery}, \bibinfo{address}{New York, NY, USA}, \bibinfo{pages}{1--15}.
\newblock
\showISBNx{978-1-4503-6708-0}
\href{https://doi.org/10.1145/3313831.3376590}{doi:\nolinkurl{10.1145/3313831.3376590}}


\bibitem[Liao et~al\mbox{.}(2022)]%
        {liaoConnectingAlgorithmicResearch2022a}
\bibfield{author}{\bibinfo{person}{Q.~Vera Liao}, \bibinfo{person}{Yunfeng Zhang}, \bibinfo{person}{Ronny Luss}, \bibinfo{person}{Finale {Doshi-Velez}}, {and} \bibinfo{person}{Amit Dhurandhar}.} \bibinfo{year}{2022}\natexlab{}.
\newblock \showarticletitle{Connecting {{Algorithmic Research}} and {{Usage Contexts}}: {{A Perspective}} of {{Contextualized Evaluation}} for {{Explainable AI}}}.
\newblock \bibinfo{journal}{\emph{Proceedings of the AAAI Conference on Human Computation and Crowdsourcing}}  \bibinfo{volume}{10} (\bibinfo{date}{Oct.} \bibinfo{year}{2022}), \bibinfo{pages}{147--159}.
\newblock
\showISSN{2769-1349}
\href{https://doi.org/10.1609/hcomp.v10i1.21995}{doi:\nolinkurl{10.1609/hcomp.v10i1.21995}}


\bibitem[Lin et~al\mbox{.}(2019)]%
        {lin2019distributional}
\bibfield{author}{\bibinfo{person}{Zichuan Lin}, \bibinfo{person}{Li Zhao}, \bibinfo{person}{Derek Yang}, \bibinfo{person}{Tao Qin}, \bibinfo{person}{Tie-Yan Liu}, {and} \bibinfo{person}{Guangwen Yang}.} \bibinfo{year}{2019}\natexlab{}.
\newblock \showarticletitle{Distributional Reward Decomposition for Reinforcement Learning}. In \bibinfo{booktitle}{\emph{Advances in Neural Information Processing Systems}}, \bibfield{editor}{\bibinfo{person}{H.~Wallach}, \bibinfo{person}{H.~Larochelle}, \bibinfo{person}{A.~Beygelzimer}, \bibinfo{person}{F.~d\textquotesingle Alch\'{e}-Buc}, \bibinfo{person}{E.~Fox}, {and} \bibinfo{person}{R.~Garnett}} (Eds.), Vol.~\bibinfo{volume}{32}. \bibinfo{publisher}{Curran Associates, Inc.}
\newblock


\bibitem[Lombrozo(2006)]%
        {lombrozo2006structure}
\bibfield{author}{\bibinfo{person}{Tania Lombrozo}.} \bibinfo{year}{2006}\natexlab{}.
\newblock \showarticletitle{The structure and function of explanations}.
\newblock \bibinfo{journal}{\emph{Trends in cognitive sciences}} \bibinfo{volume}{10}, \bibinfo{number}{10} (\bibinfo{year}{2006}), \bibinfo{pages}{464--470}.
\newblock


\bibitem[Lombrozo(2016)]%
        {lombrozo2016explanatory}
\bibfield{author}{\bibinfo{person}{Tania Lombrozo}.} \bibinfo{year}{2016}\natexlab{}.
\newblock \showarticletitle{Explanatory preferences shape learning and inference}.
\newblock \bibinfo{journal}{\emph{Trends in cognitive sciences}} \bibinfo{volume}{20}, \bibinfo{number}{10} (\bibinfo{year}{2016}), \bibinfo{pages}{748--759}.
\newblock


\bibitem[Lombrozo and Carey(2006)]%
        {lombrozo2006functional}
\bibfield{author}{\bibinfo{person}{Tania Lombrozo} {and} \bibinfo{person}{Susan Carey}.} \bibinfo{year}{2006}\natexlab{}.
\newblock \showarticletitle{Functional explanation and the function of explanation}.
\newblock \bibinfo{journal}{\emph{Cognition}} \bibinfo{volume}{99}, \bibinfo{number}{2} (\bibinfo{year}{2006}), \bibinfo{pages}{167--204}.
\newblock


\bibitem[Lombrozo and Wilkenfeld(2019)]%
        {lombrozo2019mechanistic}
\bibfield{author}{\bibinfo{person}{Tania Lombrozo} {and} \bibinfo{person}{Daniel Wilkenfeld}.} \bibinfo{year}{2019}\natexlab{}.
\newblock \showarticletitle{Mechanistic versus functional understanding}.
\newblock \bibinfo{journal}{\emph{Varieties of understanding: New perspectives from philosophy, psychology, and theology}}  \bibinfo{volume}{209} (\bibinfo{year}{2019}).
\newblock


\bibitem[Lucas et~al\mbox{.}(2014)]%
        {lucas2014child}
\bibfield{author}{\bibinfo{person}{Christopher~G Lucas}, \bibinfo{person}{Thomas~L Griffiths}, \bibinfo{person}{Fei Xu}, \bibinfo{person}{Christine Fawcett}, \bibinfo{person}{Alison Gopnik}, \bibinfo{person}{Tamar Kushnir}, \bibinfo{person}{Lori Markson}, {and} \bibinfo{person}{Jane Hu}.} \bibinfo{year}{2014}\natexlab{}.
\newblock \showarticletitle{The child as econometrician: A rational model of preference understanding in children}.
\newblock \bibinfo{journal}{\emph{PloS one}} \bibinfo{volume}{9}, \bibinfo{number}{3} (\bibinfo{year}{2014}), \bibinfo{pages}{e92160}.
\newblock


\bibitem[Lucas and Kemp(2015)]%
        {lucas2015improved}
\bibfield{author}{\bibinfo{person}{Christopher~G Lucas} {and} \bibinfo{person}{Charles Kemp}.} \bibinfo{year}{2015}\natexlab{}.
\newblock \showarticletitle{An improved probabilistic account of counterfactual reasoning.}
\newblock \bibinfo{journal}{\emph{Psychological review}} \bibinfo{volume}{122}, \bibinfo{number}{4} (\bibinfo{year}{2015}), \bibinfo{pages}{700}.
\newblock


\bibitem[Lundberg and Lee(2017)]%
        {shap}
\bibfield{author}{\bibinfo{person}{Scott~M Lundberg} {and} \bibinfo{person}{Su-In Lee}.} \bibinfo{year}{2017}\natexlab{}.
\newblock \showarticletitle{A Unified Approach to Interpreting Model Predictions}.
\newblock In \bibinfo{booktitle}{\emph{Advances in Neural Information Processing Systems 30}}, \bibfield{editor}{\bibinfo{person}{I.~Guyon}, \bibinfo{person}{U.~V. Luxburg}, \bibinfo{person}{S.~Bengio}, \bibinfo{person}{H.~Wallach}, \bibinfo{person}{R.~Fergus}, \bibinfo{person}{S.~Vishwanathan}, {and} \bibinfo{person}{R.~Garnett}} (Eds.). \bibinfo{publisher}{Curran Associates, Inc.}, \bibinfo{pages}{4765--4774}.
\newblock
\urldef\tempurl%
\url{http://papers.nips.cc/paper/7062-a-unified-approach-to-interpreting-model-predictions.pdf}
\showURL{%
\tempurl}


\bibitem[M.~Faas et~al\mbox{.}(2021)]%
        {mfaasCalibratingPedestriansTrust2021}
\bibfield{author}{\bibinfo{person}{Stefanie M.~Faas}, \bibinfo{person}{Johannes Kraus}, \bibinfo{person}{Alexander Schoenhals}, {and} \bibinfo{person}{Martin Baumann}.} \bibinfo{year}{2021}\natexlab{}.
\newblock \showarticletitle{Calibrating {{Pedestrians}}' {{Trust}} in {{Automated Vehicles}}: {{Does}} an {{Intent Display}} in an {{External HMI Support Trust Calibration}} and {{Safe Crossing Behavior}}?}. In \bibinfo{booktitle}{\emph{Proceedings of the 2021 {{CHI Conference}} on {{Human Factors}} in {{Computing Systems}}}} \emph{(\bibinfo{series}{{{CHI}} '21})}. \bibinfo{publisher}{Association for Computing Machinery}, \bibinfo{address}{New York, NY, USA}, \bibinfo{pages}{1--17}.
\newblock
\showISBNx{978-1-4503-8096-6}
\href{https://doi.org/10.1145/3411764.3445738}{doi:\nolinkurl{10.1145/3411764.3445738}}


\bibitem[Madumal et~al\mbox{.}(2018)]%
        {madumalGroundedDialogModel2018}
\bibfield{author}{\bibinfo{person}{Prashan Madumal}, \bibinfo{person}{Tim Miller}, \bibinfo{person}{Frank Vetere}, {and} \bibinfo{person}{Liz Sonenberg}.} \bibinfo{year}{2018}\natexlab{}.
\newblock \bibinfo{title}{Towards a {{Grounded Dialog Model}} for {{Explainable Artificial Intelligence}}}.
\newblock
\showeprint[arxiv]{1806.08055}~[cs]


\bibitem[Mandel and Lehman(1996)]%
        {mandel1996counterfactual}
\bibfield{author}{\bibinfo{person}{David~R Mandel} {and} \bibinfo{person}{Darrin~R Lehman}.} \bibinfo{year}{1996}\natexlab{}.
\newblock \showarticletitle{Counterfactual thinking and ascriptions of cause and preventability.}
\newblock \bibinfo{journal}{\emph{Journal of personality and social psychology}} \bibinfo{volume}{71}, \bibinfo{number}{3} (\bibinfo{year}{1996}), \bibinfo{pages}{450}.
\newblock


\bibitem[Meteyard and Davies(2020)]%
        {meteyard2020best}
\bibfield{author}{\bibinfo{person}{Lotte Meteyard} {and} \bibinfo{person}{Robert~AI Davies}.} \bibinfo{year}{2020}\natexlab{}.
\newblock \showarticletitle{Best practice guidance for linear mixed-effects models in psychological science}.
\newblock \bibinfo{journal}{\emph{Journal of Memory and Language}}  \bibinfo{volume}{112} (\bibinfo{year}{2020}), \bibinfo{pages}{104092}.
\newblock


\bibitem[Milani et~al\mbox{.}(2024)]%
        {milaniExplainableReinforcementLearning2024a}
\bibfield{author}{\bibinfo{person}{Stephanie Milani}, \bibinfo{person}{Nicholay Topin}, \bibinfo{person}{Manuela Veloso}, {and} \bibinfo{person}{Fei Fang}.} \bibinfo{year}{2024}\natexlab{}.
\newblock \showarticletitle{Explainable {{Reinforcement Learning}}: {{A Survey}} and {{Comparative Review}}}.
\newblock \bibinfo{journal}{\emph{Comput. Surveys}} \bibinfo{volume}{56}, \bibinfo{number}{7} (\bibinfo{date}{July} \bibinfo{year}{2024}), \bibinfo{pages}{1--36}.
\newblock
\showISSN{0360-0300, 1557-7341}
\href{https://doi.org/10.1145/3616864}{doi:\nolinkurl{10.1145/3616864}}


\bibitem[Miller et~al\mbox{.}(2021)]%
        {miller2021breiman}
\bibfield{author}{\bibinfo{person}{Andrew~C Miller}, \bibinfo{person}{Nicholas~J Foti}, {and} \bibinfo{person}{Emily~B Fox}.} \bibinfo{year}{2021}\natexlab{}.
\newblock \showarticletitle{Breiman's two cultures: You don't have to choose sides}.
\newblock \bibinfo{journal}{\emph{Observational Studies}} \bibinfo{volume}{7}, \bibinfo{number}{1} (\bibinfo{year}{2021}), \bibinfo{pages}{161--169}.
\newblock


\bibitem[Miller(2019)]%
        {miller2019explanation}
\bibfield{author}{\bibinfo{person}{Tim Miller}.} \bibinfo{year}{2019}\natexlab{}.
\newblock \showarticletitle{Explanation in artificial intelligence: Insights from the social sciences}.
\newblock \bibinfo{journal}{\emph{Artificial intelligence}}  \bibinfo{volume}{267} (\bibinfo{year}{2019}), \bibinfo{pages}{1--38}.
\newblock


\bibitem[Miller(2023)]%
        {millerExplainableAIDead2023}
\bibfield{author}{\bibinfo{person}{Tim Miller}.} \bibinfo{year}{2023}\natexlab{}.
\newblock \showarticletitle{Explainable {{AI}} Is {{Dead}}, {{Long Live Explainable AI}}! {{Hypothesis-driven Decision Support}} Using {{Evaluative AI}}}. In \bibinfo{booktitle}{\emph{Proceedings of the 2023 {{ACM Conference}} on {{Fairness}}, {{Accountability}}, and {{Transparency}}}} \emph{(\bibinfo{series}{{{FAccT}} '23})}. \bibinfo{publisher}{{Association for Computing Machinery}}, \bibinfo{address}{{New York, NY, USA}}, \bibinfo{pages}{333--342}.
\newblock


\bibitem[Nam et~al\mbox{.}(2023)]%
        {nam2023show}
\bibfield{author}{\bibinfo{person}{Andrew Nam}, \bibinfo{person}{Christopher Hughes}, \bibinfo{person}{Thomas Icard}, {and} \bibinfo{person}{Tobias Gerstenberg}.} \bibinfo{year}{2023}\natexlab{}.
\newblock \showarticletitle{Show and tell: Learning causal structures from observations and explanations}.
\newblock In \bibinfo{booktitle}{\emph{Proceedings of the 45th Annual Conference of the Cognitive Science Society}}.
\newblock


\bibitem[Navarre et~al\mbox{.}(2024)]%
        {navarre2024functional}
\bibfield{author}{\bibinfo{person}{Nicolas Navarre}, \bibinfo{person}{Can Konuk}, \bibinfo{person}{Neil Bramley}, {and} \bibinfo{person}{Salvador Mascarenhas}.} \bibinfo{year}{2024}\natexlab{}.
\newblock \showarticletitle{Functional rule inference from causal selection explanations}.
\newblock In \bibinfo{booktitle}{\emph{roceedings of the 46th Annual Meeting of the Cognitive Science Society}}. \bibinfo{publisher}{eScholarship}, \bibinfo{address}{Merced, CA}.
\newblock


\bibitem[Nimmo et~al\mbox{.}(2024)]%
        {nimmoUserCharacteristicsExplainable2024}
\bibfield{author}{\bibinfo{person}{Robert Nimmo}, \bibinfo{person}{Marios Constantinides}, \bibinfo{person}{Ke Zhou}, \bibinfo{person}{Daniele Quercia}, {and} \bibinfo{person}{Simone Stumpf}.} \bibinfo{year}{2024}\natexlab{}.
\newblock \showarticletitle{User {{Characteristics}} in {{Explainable AI}}: {{The Rabbit Hole}} of {{Personalization}}?}. In \bibinfo{booktitle}{\emph{Proceedings of the {{CHI Conference}} on {{Human Factors}} in {{Computing Systems}}}} \emph{(\bibinfo{series}{{{CHI}} '24})}. \bibinfo{publisher}{Association for Computing Machinery}, \bibinfo{address}{New York, NY, USA}, \bibinfo{pages}{1--13}.
\newblock
\showISBNx{9798400703300}
\href{https://doi.org/10.1145/3613904.3642352}{doi:\nolinkurl{10.1145/3613904.3642352}}


\bibitem[P{\'a}ez(2019)]%
        {paezPragmaticTurnExplainable2019a}
\bibfield{author}{\bibinfo{person}{Andr{\'e}s P{\'a}ez}.} \bibinfo{year}{2019}\natexlab{}.
\newblock \showarticletitle{The {{Pragmatic Turn}} in {{Explainable Artificial Intelligence}} ({{XAI}})}.
\newblock \bibinfo{journal}{\emph{Minds and Machines}} \bibinfo{volume}{29}, \bibinfo{number}{3} (\bibinfo{date}{Sept.} \bibinfo{year}{2019}), \bibinfo{pages}{441--459}.
\newblock


\bibitem[Pearl(2000)]%
        {pearl2000causality}
\bibfield{author}{\bibinfo{person}{Judea Pearl}.} \bibinfo{year}{2000}\natexlab{}.
\newblock \bibinfo{booktitle}{\emph{Causality}}.
\newblock \bibinfo{publisher}{Cambridge University Press}, \bibinfo{address}{Cambridge, UK}.
\newblock


\bibitem[Perez-Osorio and Wykowska(2020)]%
        {perez2020adopting}
\bibfield{author}{\bibinfo{person}{Jairo Perez-Osorio} {and} \bibinfo{person}{Agnieszka Wykowska}.} \bibinfo{year}{2020}\natexlab{}.
\newblock \showarticletitle{Adopting the intentional stance toward natural and artificial agents}.
\newblock \bibinfo{journal}{\emph{Philosophical Psychology}} \bibinfo{volume}{33}, \bibinfo{number}{3} (\bibinfo{year}{2020}), \bibinfo{pages}{369--395}.
\newblock


\bibitem[Poyiadzi et~al\mbox{.}(2020)]%
        {poyiadziFACEFeasibleActionable2020}
\bibfield{author}{\bibinfo{person}{Rafael Poyiadzi}, \bibinfo{person}{Kacper Sokol}, \bibinfo{person}{Raul {Santos-Rodriguez}}, \bibinfo{person}{Tijl De~Bie}, {and} \bibinfo{person}{Peter Flach}.} \bibinfo{year}{2020}\natexlab{}.
\newblock \showarticletitle{{{FACE}}: {{Feasible}} and {{Actionable Counterfactual Explanations}}}. In \bibinfo{booktitle}{\emph{Proceedings of the {{AAAI}}/{{ACM Conference}} on {{AI}}, {{Ethics}}, and {{Society}}}} \emph{(\bibinfo{series}{{{AIES}} '20})}. \bibinfo{publisher}{Association for Computing Machinery}, \bibinfo{address}{New York, NY, USA}, \bibinfo{pages}{344--350}.
\newblock
\showISBNx{978-1-4503-7110-0}
\href{https://doi.org/10.1145/3375627.3375850}{doi:\nolinkurl{10.1145/3375627.3375850}}


\bibitem[Quillien and Barlev(2022)]%
        {quillien2022causal}
\bibfield{author}{\bibinfo{person}{Tadeg Quillien} {and} \bibinfo{person}{Michael Barlev}.} \bibinfo{year}{2022}\natexlab{}.
\newblock \showarticletitle{Causal Judgment in the Wild: Evidence from the 2020 U.S. Presidential Election}.
\newblock \bibinfo{journal}{\emph{Cognitive Science}} \bibinfo{volume}{46}, \bibinfo{number}{2} (\bibinfo{year}{2022}), \bibinfo{pages}{e13101}.
\newblock
\href{https://doi.org/10.1111/cogs.13101}{doi:\nolinkurl{10.1111/cogs.13101}}


\bibitem[Quillien and German(2021)]%
        {quillien2021simple}
\bibfield{author}{\bibinfo{person}{Tadeg Quillien} {and} \bibinfo{person}{Tamsin~C German}.} \bibinfo{year}{2021}\natexlab{}.
\newblock \showarticletitle{A simple definition of ‘intentionally’}.
\newblock \bibinfo{journal}{\emph{Cognition}}  \bibinfo{volume}{214} (\bibinfo{year}{2021}), \bibinfo{pages}{104806}.
\newblock


\bibitem[Quillien and Lucas(2023)]%
        {quillien2023counterfactuals}
\bibfield{author}{\bibinfo{person}{Tadeg Quillien} {and} \bibinfo{person}{Christopher~G Lucas}.} \bibinfo{year}{2023}\natexlab{}.
\newblock \showarticletitle{Counterfactuals and the logic of causal selection}.
\newblock \bibinfo{journal}{\emph{Psychological Review}}  \bibinfo{volume}{131} (\bibinfo{year}{2023}), \bibinfo{pages}{1208–--1234}.
\newblock
Issue 5.
\href{https://doi.org/10.1037/rev0000428}{doi:\nolinkurl{10.1037/rev0000428}}


\bibitem[Rudin(2019)]%
        {rudinStopExplainingBlack2019}
\bibfield{author}{\bibinfo{person}{Cynthia Rudin}.} \bibinfo{year}{2019}\natexlab{}.
\newblock \showarticletitle{Stop Explaining Black Box Machine Learning Models for High Stakes Decisions and Use Interpretable Models Instead}.
\newblock \bibinfo{journal}{\emph{Nature Machine Intelligence}} \bibinfo{volume}{1}, \bibinfo{number}{5} (\bibinfo{date}{May} \bibinfo{year}{2019}), \bibinfo{pages}{206--215}.
\newblock
\showISSN{2522-5839}
\href{https://doi.org/10.1038/s42256-019-0048-x}{doi:\nolinkurl{10.1038/s42256-019-0048-x}}


\bibitem[Saeed and Omlin(2023)]%
        {saeedExplainableAIXAI2023}
\bibfield{author}{\bibinfo{person}{Waddah Saeed} {and} \bibinfo{person}{Christian Omlin}.} \bibinfo{year}{2023}\natexlab{}.
\newblock \showarticletitle{Explainable {{AI}} ({{XAI}}): {{A}} Systematic Meta-Survey of Current Challenges and Future Opportunities}.
\newblock \bibinfo{journal}{\emph{Knowledge-Based Systems}}  \bibinfo{volume}{263} (\bibinfo{date}{March} \bibinfo{year}{2023}), \bibinfo{pages}{110273}.
\newblock
\showISSN{0950-7051}
\href{https://doi.org/10.1016/j.knosys.2023.110273}{doi:\nolinkurl{10.1016/j.knosys.2023.110273}}


\bibitem[Schwalbe and Finzel(2023)]%
        {schwalbe2023comprehensive}
\bibfield{author}{\bibinfo{person}{Gesina Schwalbe} {and} \bibinfo{person}{Bettina Finzel}.} \bibinfo{year}{2023}\natexlab{}.
\newblock \showarticletitle{A comprehensive taxonomy for explainable artificial intelligence: a systematic survey of surveys on methods and concepts}.
\newblock \bibinfo{journal}{\emph{Data Mining and Knowledge Discovery}} \bibinfo{volume}{38}, \bibinfo{number}{5} (\bibinfo{date}{Jan.} \bibinfo{year}{2023}), \bibinfo{pages}{3043–3101}.
\newblock
\showISSN{1384-5810}
\href{https://doi.org/10.1007/s10618-022-00867-8}{doi:\nolinkurl{10.1007/s10618-022-00867-8}}


\bibitem[Septon et~al\mbox{.}(2023)]%
        {septon2023integrating}
\bibfield{author}{\bibinfo{person}{Yael Septon}, \bibinfo{person}{Tobias Huber}, \bibinfo{person}{Elisabeth Andr\'{e}}, {and} \bibinfo{person}{Ofra Amir}.} \bibinfo{year}{2023}\natexlab{}.
\newblock \showarticletitle{Integrating Policy Summaries with Reward Decomposition for Explaining Reinforcement Learning Agents}. In \bibinfo{booktitle}{\emph{Advances in Practical Applications of Agents, Multi-Agent Systems, and Cognitive Mimetics. The PAAMS Collection: 21st International Conference, PAAMS 2023, Guimar\~{a}es, Portugal, July 12–14, 2023, Proceedings}} (Guimaraes, Portugal). \bibinfo{publisher}{Springer-Verlag}, \bibinfo{address}{Berlin, Heidelberg}, \bibinfo{pages}{320–332}.
\newblock
\showISBNx{978-3-031-37615-3}
\href{https://doi.org/10.1007/978-3-031-37616-0_27}{doi:\nolinkurl{10.1007/978-3-031-37616-0_27}}


\bibitem[Slack et~al\mbox{.}(2023)]%
        {slackExplainingMachineLearning2023}
\bibfield{author}{\bibinfo{person}{Dylan Slack}, \bibinfo{person}{Satyapriya Krishna}, \bibinfo{person}{Himabindu Lakkaraju}, {and} \bibinfo{person}{Sameer Singh}.} \bibinfo{year}{2023}\natexlab{}.
\newblock \showarticletitle{Explaining Machine Learning Models with Interactive Natural Language Conversations Using {{TalkToModel}}}.
\newblock \bibinfo{journal}{\emph{Nature Machine Intelligence}} \bibinfo{volume}{5}, \bibinfo{number}{8} (\bibinfo{date}{Aug.} \bibinfo{year}{2023}), \bibinfo{pages}{873--883}.
\newblock


\bibitem[Speith(2022)]%
        {speithReviewTaxonomiesExplainable2022}
\bibfield{author}{\bibinfo{person}{Timo Speith}.} \bibinfo{year}{2022}\natexlab{}.
\newblock \showarticletitle{A {{Review}} of {{Taxonomies}} of {{Explainable Artificial Intelligence}} ({{XAI}}) {{Methods}}}. In \bibinfo{booktitle}{\emph{2022 {{ACM Conference}} on {{Fairness}}, {{Accountability}}, and {{Transparency}}}} \emph{(\bibinfo{series}{{{FAccT}} '22})}. \bibinfo{publisher}{Association for Computing Machinery}, \bibinfo{address}{New York, NY, USA}, \bibinfo{pages}{2239--2250}.
\newblock


\bibitem[Sreedharan et~al\mbox{.}(2021)]%
        {sreedharanFoundationsExplanationsModel2021}
\bibfield{author}{\bibinfo{person}{Sarath Sreedharan}, \bibinfo{person}{Tathagata Chakraborti}, {and} \bibinfo{person}{Subbarao Kambhampati}.} \bibinfo{year}{2021}\natexlab{}.
\newblock \showarticletitle{Foundations of Explanations as Model Reconciliation}.
\newblock \bibinfo{journal}{\emph{Artificial Intelligence}}  \bibinfo{volume}{301} (\bibinfo{date}{Dec.} \bibinfo{year}{2021}), \bibinfo{pages}{103558}.
\newblock
\showISSN{0004-3702}
\href{https://doi.org/10.1016/j.artint.2021.103558}{doi:\nolinkurl{10.1016/j.artint.2021.103558}}


\bibitem[Stepin et~al\mbox{.}(2021)]%
        {stepinSurveyContrastiveCounterfactual2021}
\bibfield{author}{\bibinfo{person}{Ilia Stepin}, \bibinfo{person}{Jose~M. Alonso}, \bibinfo{person}{Alejandro Catala}, {and} \bibinfo{person}{Mart{\'i}n {Pereira-Fari{\~n}a}}.} \bibinfo{year}{2021}\natexlab{}.
\newblock \showarticletitle{A {{Survey}} of {{Contrastive}} and {{Counterfactual Explanation Generation Methods}} for {{Explainable Artificial Intelligence}}}.
\newblock \bibinfo{journal}{\emph{IEEE Access}}  \bibinfo{volume}{9} (\bibinfo{year}{2021}), \bibinfo{pages}{11974--12001}.
\newblock


\bibitem[Sulik et~al\mbox{.}(2023)]%
        {sulik2023explanations}
\bibfield{author}{\bibinfo{person}{Justin Sulik}, \bibinfo{person}{Jeroen van Paridon}, {and} \bibinfo{person}{Gary Lupyan}.} \bibinfo{year}{2023}\natexlab{}.
\newblock \showarticletitle{Explanations in the wild}.
\newblock \bibinfo{journal}{\emph{Cognition}}  \bibinfo{volume}{237} (\bibinfo{year}{2023}), \bibinfo{pages}{105464}.
\newblock


\bibitem[Taylor and Taylor(2021)]%
        {taylorArtificialCognitionHow2021}
\bibfield{author}{\bibinfo{person}{J.~Eric~T. Taylor} {and} \bibinfo{person}{Graham~W. Taylor}.} \bibinfo{year}{2021}\natexlab{}.
\newblock \showarticletitle{Artificial Cognition: {{How}} Experimental Psychology Can Help Generate Explainable Artificial Intelligence}.
\newblock \bibinfo{journal}{\emph{Psychonomic Bulletin \& Review}} \bibinfo{volume}{28}, \bibinfo{number}{2} (\bibinfo{date}{April} \bibinfo{year}{2021}), \bibinfo{pages}{454--475}.
\newblock


\bibitem[Ustun et~al\mbox{.}(2019)]%
        {ustunActionableRecourseLinear2019}
\bibfield{author}{\bibinfo{person}{Berk Ustun}, \bibinfo{person}{Alexander Spangher}, {and} \bibinfo{person}{Yang Liu}.} \bibinfo{year}{2019}\natexlab{}.
\newblock \showarticletitle{Actionable Recourse in Linear Classification}. In \bibinfo{booktitle}{\emph{Proceedings of the Conference on Fairness, Accountability, and Transparency}} (Atlanta, GA, USA) \emph{(\bibinfo{series}{FAT* '19})}. \bibinfo{publisher}{Association for Computing Machinery}, \bibinfo{address}{New York, NY, USA}, \bibinfo{pages}{10–19}.
\newblock
\showISBNx{9781450361255}
\href{https://doi.org/10.1145/3287560.3287566}{doi:\nolinkurl{10.1145/3287560.3287566}}


\bibitem[van~der Waa et~al\mbox{.}(2021)]%
        {van2021evaluating}
\bibfield{author}{\bibinfo{person}{Jasper van~der Waa}, \bibinfo{person}{Elisabeth Nieuwburg}, \bibinfo{person}{Anita Cremers}, {and} \bibinfo{person}{Mark Neerincx}.} \bibinfo{year}{2021}\natexlab{}.
\newblock \showarticletitle{Evaluating XAI: A comparison of rule-based and example-based explanations}.
\newblock \bibinfo{journal}{\emph{Artificial intelligence}}  \bibinfo{volume}{291} (\bibinfo{year}{2021}), \bibinfo{pages}{103404}.
\newblock


\bibitem[Wang et~al\mbox{.}(2019)]%
        {wangDesigningTheoryDrivenUserCentric2019a}
\bibfield{author}{\bibinfo{person}{Danding Wang}, \bibinfo{person}{Qian Yang}, \bibinfo{person}{Ashraf Abdul}, {and} \bibinfo{person}{Brian~Y. Lim}.} \bibinfo{year}{2019}\natexlab{}.
\newblock \showarticletitle{Designing {{Theory-Driven User-Centric Explainable AI}}}. In \bibinfo{booktitle}{\emph{Proceedings of the 2019 {{CHI Conference}} on {{Human Factors}} in {{Computing Systems}}}}. \bibinfo{publisher}{ACM}, \bibinfo{address}{Glasgow Scotland Uk}, \bibinfo{pages}{1--15}.
\newblock
\showISBNx{978-1-4503-5970-2}
\href{https://doi.org/10.1145/3290605.3300831}{doi:\nolinkurl{10.1145/3290605.3300831}}


\bibitem[Warren et~al\mbox{.}(2023)]%
        {warren2023categorical}
\bibfield{author}{\bibinfo{person}{Greta Warren}, \bibinfo{person}{Ruth M.~J. Byrne}, {and} \bibinfo{person}{Mark~T. Keane}.} \bibinfo{year}{2023}\natexlab{}.
\newblock \showarticletitle{Categorical and Continuous Features in Counterfactual Explanations of AI Systems}. In \bibinfo{booktitle}{\emph{Proceedings of the 28th International Conference on Intelligent User Interfaces}} (Sydney, NSW, Australia) \emph{(\bibinfo{series}{IUI '23})}. \bibinfo{publisher}{Association for Computing Machinery}, \bibinfo{address}{New York, NY, USA}, \bibinfo{pages}{171–187}.
\newblock
\showISBNx{9798400701061}
\href{https://doi.org/10.1145/3581641.3584090}{doi:\nolinkurl{10.1145/3581641.3584090}}


\bibitem[Warren et~al\mbox{.}(2024)]%
        {warren2024explaining}
\bibfield{author}{\bibinfo{person}{Greta Warren}, \bibinfo{person}{Eoin Delaney}, \bibinfo{person}{Christophe Gu{\'{e}}ret}, {and} \bibinfo{person}{Mark~T. Keane}.} \bibinfo{year}{2024}\natexlab{}.
\newblock \showarticletitle{Explaining Multiple Instances Counterfactually: User Tests of Group-Counterfactuals for {XAI}}. In \bibinfo{booktitle}{\emph{Case-Based Reasoning Research and Development - 32nd International Conference, {ICCBR} 2024, Merida, Mexico, July 1-4, 2024, Proceedings}} \emph{(\bibinfo{series}{Lecture Notes in Computer Science}, Vol.~\bibinfo{volume}{14775})}, \bibfield{editor}{\bibinfo{person}{Juan~A. Recio{-}Garc{\'{\i}}a}, \bibinfo{person}{Mauricio~Gabriel Orozco{-}del{-}Castillo}, {and} \bibinfo{person}{Derek Bridge}} (Eds.). \bibinfo{publisher}{Springer}, \bibinfo{address}{New York, NY}, \bibinfo{pages}{206--222}.
\newblock
\href{https://doi.org/10.1007/978-3-031-63646-2\_14}{doi:\nolinkurl{10.1007/978-3-031-63646-2\_14}}


\bibitem[Wiegand et~al\mbox{.}(2020)]%
        {wiegandExplanationThatExploring2020}
\bibfield{author}{\bibinfo{person}{Gesa Wiegand}, \bibinfo{person}{Malin Eiband}, \bibinfo{person}{Maximilian Haubelt}, {and} \bibinfo{person}{Heinrich Hussmann}.} \bibinfo{year}{2020}\natexlab{}.
\newblock \showarticletitle{"{{I}}'d like an {{Explanation}} for {{That}}!" {{Exploring Reactions}} to {{Unexpected Autonomous Driving}}}. In \bibinfo{booktitle}{\emph{22nd {{International Conference}} on {{Human-Computer Interaction}} with {{Mobile Devices}} and {{Services}}}} \emph{(\bibinfo{series}{{{MobileHCI}} '20})}. \bibinfo{publisher}{Association for Computing Machinery}, \bibinfo{address}{New York, NY, USA}, \bibinfo{pages}{1--11}.
\newblock
\showISBNx{978-1-4503-7516-0}
\href{https://doi.org/10.1145/3379503.3403554}{doi:\nolinkurl{10.1145/3379503.3403554}}


\bibitem[Woodward(2004)]%
        {woodward2003making}
\bibfield{author}{\bibinfo{person}{James Woodward}.} \bibinfo{year}{2004}\natexlab{}.
\newblock \bibinfo{booktitle}{\emph{Making Things Happen: A Theory of Causal Explanation}}.
\newblock \bibinfo{publisher}{Oxford University Press}, \bibinfo{address}{Oxford, UK}.
\newblock
\showISBNx{9780195155273}
\href{https://doi.org/10.1093/0195155270.001.0001}{doi:\nolinkurl{10.1093/0195155270.001.0001}}


\bibitem[Yang et~al\mbox{.}(2022)]%
        {yangPsychologicalTheoryExplainability2022}
\bibfield{author}{\bibinfo{person}{Scott Cheng-Hsin Yang}, \bibinfo{person}{Nils Erik~Tomas Folke}, {and} \bibinfo{person}{Patrick Shafto}.} \bibinfo{year}{2022}\natexlab{}.
\newblock \showarticletitle{A {{Psychological Theory}} of {{Explainability}}}. In \bibinfo{booktitle}{\emph{Proceedings of the 39th {{International Conference}} on {{Machine Learning}}}}. \bibinfo{publisher}{{PMLR}}, \bibinfo{pages}{25007--25021}.
\newblock


\bibitem[Yuan et~al\mbox{.}(2023)]%
        {yuanContextualizingUserPerceptions2023}
\bibfield{author}{\bibinfo{person}{Chien Wen~(Tina) Yuan}, \bibinfo{person}{Nanyi Bi}, \bibinfo{person}{Ya-Fang Lin}, {and} \bibinfo{person}{Yuen-Hsien Tseng}.} \bibinfo{year}{2023}\natexlab{}.
\newblock \showarticletitle{Contextualizing {{User Perceptions}} about {{Biases}} for {{Human-Centered Explainable Artificial Intelligence}}}. In \bibinfo{booktitle}{\emph{Proceedings of the 2023 {{CHI Conference}} on {{Human Factors}} in {{Computing Systems}}}} \emph{(\bibinfo{series}{{{CHI}} '23})}. \bibinfo{publisher}{Association for Computing Machinery}, \bibinfo{address}{New York, NY, USA}, \bibinfo{pages}{1--15}.
\newblock
\showISBNx{978-1-4503-9421-5}
\href{https://doi.org/10.1145/3544548.3580945}{doi:\nolinkurl{10.1145/3544548.3580945}}


\bibitem[Zednik(2021)]%
        {zednikSolvingBlackBox2021}
\bibfield{author}{\bibinfo{person}{Carlos Zednik}.} \bibinfo{year}{2021}\natexlab{}.
\newblock \showarticletitle{Solving the {{Black Box Problem}}: {{A Normative Framework}} for {{Explainable Artificial Intelligence}}}.
\newblock \bibinfo{journal}{\emph{Philosophy \& Technology}} \bibinfo{volume}{34}, \bibinfo{number}{2} (\bibinfo{date}{June} \bibinfo{year}{2021}), \bibinfo{pages}{265--288}.
\newblock


\bibitem[Zemla et~al\mbox{.}(2017)]%
        {zemla2017evaluating}
\bibfield{author}{\bibinfo{person}{Jeffrey~C Zemla}, \bibinfo{person}{Steven Sloman}, \bibinfo{person}{Christos Bechlivanidis}, {and} \bibinfo{person}{David~A Lagnado}.} \bibinfo{year}{2017}\natexlab{}.
\newblock \showarticletitle{Evaluating everyday explanations}.
\newblock \bibinfo{journal}{\emph{Psychonomic bulletin \& review}}  \bibinfo{volume}{24} (\bibinfo{year}{2017}), \bibinfo{pages}{1488--1500}.
\newblock


\bibitem[Zerilli(2022)]%
        {zerilli2022explaining}
\bibfield{author}{\bibinfo{person}{John Zerilli}.} \bibinfo{year}{2022}\natexlab{}.
\newblock \showarticletitle{Explaining machine learning decisions}.
\newblock \bibinfo{journal}{\emph{Philosophy of Science}} \bibinfo{volume}{89}, \bibinfo{number}{1} (\bibinfo{year}{2022}), \bibinfo{pages}{1--19}.
\newblock


\bibitem[Zhang et~al\mbox{.}(2017)]%
        {zhang2017plan}
\bibfield{author}{\bibinfo{person}{Yu Zhang}, \bibinfo{person}{Sarath Sreedharan}, \bibinfo{person}{Anagha Kulkarni}, \bibinfo{person}{Tathagata Chakraborti}, \bibinfo{person}{Hankz~Hankui Zhuo}, {and} \bibinfo{person}{Subbarao Kambhampati}.} \bibinfo{year}{2017}\natexlab{}.
\newblock \showarticletitle{Plan explicability and predictability for robot task planning}. In \bibinfo{booktitle}{\emph{2017 IEEE International Conference on Robotics and Automation (ICRA)}} (Singapore). \bibinfo{publisher}{IEEE}, \bibinfo{address}{New York, NY}, \bibinfo{pages}{1313--1320}.
\newblock
\href{https://doi.org/10.1109/ICRA.2017.7989155}{doi:\nolinkurl{10.1109/ICRA.2017.7989155}}


\end{thebibliography}

\end{document}